\documentclass[10pt, final]{IEEEtran}

\usepackage{amsmath,stmaryrd,acronym,amssymb,amsthm,dsfont,graphicx}
\graphicspath{{graphics/}}

\acrodef{AEP}{Asymptotic Equipartition Property}
\acrodef{AoA}{Angle of Arrival}
\acrodef{AWGN}{Additive White Gaussian Noise}
\acrodef{BER}{Bit-Error-Rate}
\acrodef{BEC}{Binary Erasure Channel}
\acrodef{BPSK}{Binary Phase-Shift Keying}
\acrodef{BSC}{Binary Symmetric Channel}
\acrodef{CDF}[CDF]{Cumulative Distribution Function}
\acrodef{CLT}[CLT]{Central Limit Theorem}
\acrodef{CSI}[CSI]{Channel State Information}
\acrodef{DMC}[DMC]{Discrete Memoryless Channel}
\acrodef{DMS}[DMS]{Discrete Memoryless Source}
\acrodef{iid}[i.i.d.]{independent and identically distributed}
\acrodef{lhs}[l.h.s.]{left-hand-side}
\acrodef{rhs}[r.h.s.]{right-hand-side}
\acrodef{LPD}[LPD]{Low Probability of Detection}
\acrodef{LDPC}[LDPC]{Low-Density Parity-Check}
\acrodef{MAC}[MAC]{multiple-access channel}
\acrodef{MIMO}[MIMO]{Multiple-Input Multiple-Output}
\acrodef{MISO}{Multiple-Input Single-Output}
\acrodef{PDF}[PDF]{Probability Distribution Function}
\acrodef{PMF}[PMF]{Probability Mass Function}
\acrodef{PPM}[PPM]{Pulse Position Modulation}
\acrodef{PSD}{Power Spectral Density}
\acrodef{QPSK}{Quadrature Phase-Shift Keying}
\acrodef{SIMO}{Single-Input Multiple-Output}
\acrodef{SNR}{Signal-to-Noise Ratio}
\acrodef{wrt}[w.r.t.]{with respect to}
\acrodef{WSS}{Wide Sense Stationary}
\acrodef{AVC}{Arbitrarily Varying Channel}

\DeclareMathAlphabet{\eurm}{U}{eur}{m}{n}
\DeclareMathAlphabet{\mathbsf}{OT1}{cmss}{bx}{n}
\DeclareMathAlphabet{\mathssf}{OT1}{cmss}{m}{sl}
\DeclareMathAlphabet{\mathcsf}{OT1}{cmss}{sbc}{n}

\DeclareSymbolFont{bsfletters}{OT1}{cmss}{bx}{n}  
\DeclareSymbolFont{ssfletters}{OT1}{cmss}{m}{n}
\DeclareMathSymbol{\bsfGamma}{0}{bsfletters}{'000}
\DeclareMathSymbol{\ssfGamma}{0}{ssfletters}{'000}
\DeclareMathSymbol{\bsfDelta}{0}{bsfletters}{'001}
\DeclareMathSymbol{\ssfDelta}{0}{ssfletters}{'001}
\DeclareMathSymbol{\bsfTheta}{0}{bsfletters}{'002}
\DeclareMathSymbol{\ssfTheta}{0}{ssfletters}{'002}
\DeclareMathSymbol{\bsfLambda}{0}{bsfletters}{'003}
\DeclareMathSymbol{\ssfLambda}{0}{ssfletters}{'003}
\DeclareMathSymbol{\bsfXi}{0}{bsfletters}{'004}
\DeclareMathSymbol{\ssfXi}{0}{ssfletters}{'004}
\DeclareMathSymbol{\bsfPi}{0}{bsfletters}{'005}
\DeclareMathSymbol{\ssfPi}{0}{ssfletters}{'005}
\DeclareMathSymbol{\bsfSigma}{0}{bsfletters}{'006}
\DeclareMathSymbol{\ssfSigma}{0}{ssfletters}{'006}
\DeclareMathSymbol{\bsfUpsilon}{0}{bsfletters}{'007}
\DeclareMathSymbol{\ssfUpsilon}{0}{ssfletters}{'007}
\DeclareMathSymbol{\bsfPhi}{0}{bsfletters}{'010}
\DeclareMathSymbol{\ssfPhi}{0}{ssfletters}{'010}
\DeclareMathSymbol{\bsfPsi}{0}{bsfletters}{'011}
\DeclareMathSymbol{\ssfPsi}{0}{ssfletters}{'011}
\DeclareMathSymbol{\bsfOmega}{0}{bsfletters}{'012}
\DeclareMathSymbol{\ssfOmega}{0}{ssfletters}{'012}

\newcommand{\calC}{{\mathcal{C}}}

\newcommand{\calE}{{\mathcal{E}}}
\newcommand{\calF}{{\mathcal{F}}}

\newcommand{\calJ}{{\mathcal{J}}}

\newcommand{\calP}{{\mathcal{P}}}

\newcommand{\calR}{{\mathcal{R}}}
\newcommand{\calT}{{\mathcal{T}}}
\newcommand{\calS}{{\mathcal{S}}}

\newcommand{\calX}{{\mathcal{X}}}
\newcommand{\calY}{{\mathcal{Y}}}

\newcommand{\calZ}{{\mathcal{Z}}}
\newcommand{\modfirst}[1]{{#1}}

\newcommand{\E}[2][]{{\mathbb{E}_{#1}}{\left(#2\right)}}       
\renewcommand{\P}[2][]{{\mathbb{P}_{#1}}{\left(#2\right)}}
\newcommand{\Var}[1]{{\text{\textnormal{Var}}{\left(#1\right)}}}       
\newcommand{\D}[2]{{{\mathbb{D}}\!\left({#1\Vert#2}\right)}}

\newcommand{\V}[1]{{{\mathbb{V}}\!\left(#1\right)}}

\newcommand{\avgI}[1]{{{\mathbb{I}}\!\left(#1\right)}}
\newcommand{\avgH}[1]{{\mathbb{H}}\!\left(#1\right)}

\newcommand{\Hb}[1]{{\mathbb{H}_b}\left(#1\right)}

\newcommand{\wt}[1]{\ensuremath{\textnormal{wt}(#1)}}

\newcommand{\card}[1]{\ensuremath{\left|{#1}\right|}}           
\newcommand{\eqdef}{\ensuremath{\triangleq}}                    
\newcommand{\intseq}[2]{\ensuremath{\llbracket{#1},{#2}\rrbracket}}  
\newcommand{\indic}[1]{\ensuremath{\mathds{1}\!\left\{#1\right\}}}

\renewcommand{\leq}{\leqslant}
\renewcommand{\geq}{\geqslant}

\newcommand{\proddist}{%
  \mathchoice{\raisebox{1pt}{$\displaystyle\otimes$}}
             {\raisebox{1pt}{$\otimes$}}
             {\raisebox{0.5pt}{\scalebox{0.7}{$\scriptstyle\otimes$}}}
             {\raisebox{0.4pt}{\scalebox{0.6}{$\scriptscriptstyle\otimes$}}}}
\newcommand{\pn}{{\proddist n}}

\usepackage[textwidth=0.68in,textsize=footnotesize,disable]{todonotes}
\presetkeys{todonotes}{color=redArcom, linecolor=redArcom,bordercolor=redArcom}{}

\newtheorem{theorem}{Theorem}
\newtheorem{remark}{Remark}
\newtheorem{definition}{Definition}
\newtheorem{lemma}{Lemma}
\newtheorem{corollary}{Corollary}

\acrodef{ROC}[ROC]{Receiver Operation Characteristic}
\acrodef{PPM}[PPM]{Pulse-Position Modulation}

\newcommand{\pr}[1]{{\left(#1\right)}}

\begin{document}

\title{Covert Secret Key Generation with an Active Warden}
\author{Mehrdad Tahmasbi and Matthieu R. Bloch}

\maketitle

\begin{abstract}
\modfirst{We investigate the problem of covert and secret key generation over a state-dependent discrete memoryless channel with one-way public discussion in which an adversary, the warden, may arbitrarily choose the channel state. 
We develop an adaptive protocol that, under conditions that we explicitly specify, not only allows the transmitter and the legitimate receiver to exchange a secret key but also conceals from the active warden whether the protocol is being run. When specialized to passive adversaries that do not control the channel state, we partially characterize the covert secret key capacity.  
In particular, the covert secret key capacity is sometimes equal to the covert capacity of the channel, so that secrecy comes ``for free.''}
\end{abstract}
\section{Introduction}
\label{sec:introduction}

\modfirst{Following the early results of Ahlswede and Csisz\'ar~\cite{Ahlswede1993} and Maurer~\cite{Maurer1993}, secret key generation from correlated observations using an authenticated public channel has attracted significant attention, especially in the context of wireless channels \cite{Mukherjee2014, Ye2010}. We investigate here the problem of \emph{covert} secret key generation, in which legitimate parties must not only agree on a common secret key but also keep the key generation protocol undetectable by an adversary, referred to as the \emph{warden}.

Our work builds upon recent results on covert~\cite{Bash2013,Che2013,Bloch2016a,Wang2016b} and stealth~\cite{Hou2014, Wu2015} communications, which have characterized how many information bits can be transmitted reliably over noisy channels while escaping detection by a warden. In particular, the number of reliable and covert bits that can be transmitted in $n$ channel uses is limited by a \emph{square root law} to $O(\sqrt{n})$~\cite{Bash2013}. When the warden's channel is of higher quality, in a sense precisely defined in~\cite{Bloch2016a}, covert communication is possible but at the expense of using a shared secret key between legitimate parties. Notice that this implicitly requires the existence of a secret key exchange mechanism that does improve the detection capability of the warden. Our work also capitalizes on efforts to consider state-dependent models for covert communication such as \cite{shahzad2017covert}, in which covert rates are identified for a fading channel with randomly varying states only statistically known to the legitimate users, or \cite{lee2018covert}, in which the covert capacity is characterized for a state-dependent channels with causal or non-causal channel knowledge at the transmitter. Another relevant work is \cite{sobers2017covert}, in which the warden's uncertainty about the channel is shown to allows the circumvention of the square root law. Finally, covert communication over adversarial channels, in which the warden flips a certain fraction of the transmitted bits, has been investigated in~\cite{zhang2018covert}; covert communication is shown to be possible for all warden's states if the legitimate users have access to enough shared secret key, which again prompts the question of how to covertly generate such a secret key.

Our work partially addresses the question by showing that, under conditions that we shall precisely specify, covert and secret key generation is possible even with an active warden. Specifically, the results reported here extend our preliminary results restricted to a passive warden~\cite{Tahmasbi2017} to a model with an active warden who can arbitrarily vary the channel state, except when no information is sent on the main channel. While this restriction arises from the technicalities in our proofs, it is justified in certain practical scenarios. Specifically, for wireless channels in which the action of the warden corresponds to tampering with the gain of the legitimate receiver, the gain has no effect when no signal is transmitted because fading acts as a multiplicative coefficient.  

As in most results on secret key generation, the presence of an authenticated public communication is pivotal in our coding scheme to enable covert secret key generation and therefore covert communication for channels over which a secret key is required~\cite{Bloch2016a}. To avoid improving the warden's detection ability, we impose a probability distribution on the public communication and make certain that the warden cannot detect the communication with any test jointly performed on the observations of the noisy channel and of the public channel. This model relates to \emph{stealth secret key generation} from a source model~\cite{Lin2017}; however, stealth is a less stringent requirement than covertness, so that our results are of a different nature and exploit different proof techniques to characterize the covert secret key capacity. We emphasize that our approach differs from previous studies on two accounts. First, we neither impose any limit on the warden's actions nor consider any statistical model for the channel states, so that the warden may take any action and our coding scheme remains reliable and covert for all possible state sequences. Second, the covert throughput is adapted to the warden's actions, i.e., legitimate parties decide how many bits are extracted based on the quality of the channels, and we use ideas for estimation that we introduced in~\cite{tahmasbi2017learning} in the context of learning over wiretap channels.}



The remainder of the paper is organized as follows. In Section~\ref{sec:problem-formulation}, we formally introduce our model for covert secret key generation. In Section~\ref{sec:main-results-passive} and Section~\ref{sec:main-results-active}, we develop our results on covert secret key generation for passive and active models, respectively.

\section{Notation and Problem Formulation}
\label{sec:problem-formulation}

\subsection{Notation}
We denote random variables by uppercase letters (e.g., $X$), their realizations by lowercase letters (e.g., $x$), sets by calligraphic letters (e.g., $\calX$), and vectors by bold face letters (e.g., $\mathbf{x}$). For $\mathbf{x} = (x_1, \cdots, x_n) \in \calX^n$ and $a\in \calX$,  let $N(\mathbf{x}|a)\eqdef |\{i: x_i = a\}|$. For $\mathbf{x}\in\{0, 1\}^n$, let $\wt{\mathbf{x}} \eqdef N(\mathbf{x}|1)$ and $\alpha(\mathbf{x})\eqdef \frac{\wt{\mathbf{x}}}{n}$. If $P_X$ is a \ac{PMF} over $\calX$,  let $\calT_{P_X} \eqdef \{\mathbf{x} \in\calX^n: \text{for all } a \in \calX:~N(\mathbf{x}|a) = P(a) n\}$. We denote by $\calP_n(\calX)$ the set of all \acp{PMF} $P_X$ for which $\calT_{P_X} \neq \emptyset$ and by $\calP_n(\calX|\calY)$ the set of all conditional \acp{PMF} $P_{X|Y}$  for which there exists a joint \ac{PMF} $P_{XY}$ such that $P_{X|Y}= \frac{P_{XY}}{P_Y}$ and  $\calT_{P_{XY}} \neq 0$. For $\mathbf{x} \in \calX^n$ and a conditional \ac{PMF} $P_{Y|X}$, we also define $\calT_{P_{Y|X}}(\mathbf{x})\eqdef \{\mathbf{y}\in \calY^n:\text{for all }a\in \calX, b\in \calY:  N(\mathbf{x}, \mathbf{y}|a, b) = P_{Y|X}(b|a) N(\mathbf{x}|a)\}$. For three discrete random variables $(X, Y, Z)$ with joint \ac{PMF} $P_{XYZ}$, we define
\begin{align}
P_{X|YZ} \circ P_{Z} &\eqdef \sum_z P_{X|YZ=z}P_Z(z) \eqdef P_{X|Y},\\
P_{Z|Y} \times P_{X|YZ} &\eqdef P_{Z|Y} P_{X|YZ} \eqdef P_{XZ|Y},\\
I(P_X, P_{Y|X}) &\eqdef I(P_{XY}) \eqdef \avgI{X;Y}.
\end{align}
For two sequences $\mathbf{x} \in \calX^n$ and $\mathbf{y} \in \calY^n$ such that  $(\mathbf{x}, \mathbf{y}) \in \calT_{P_{XY}}$, we define $I(\mathbf{x}\wedge \mathbf{y}) \eqdef I(P_{XY}).$ For two integers $a$ and $b$ such that $a\leq b$, we denote the set $\{a, a+1, \cdots, b - 1, b\}$ by $\intseq{a}{b}$. If $a>b$, then $\intseq{a}{b}\eqdef \emptyset$. Throughout the paper, we measure the information in bits and $\log(\cdot)$ should be understood to be base $2$; we use $\ln(\cdot)$ for the logarithm base $e$.  \modfirst{We denote by $\P[P]{\cdot}$ the probability measure induced by a \ac{PMF} $P$. $P^{\mathrm{unif}}_X$ is the uniform probability distribution over $\calX$.}
\subsection{Problem Formulation}
\begin{figure}[h]
  \centering
  \includegraphics[width=\linewidth]{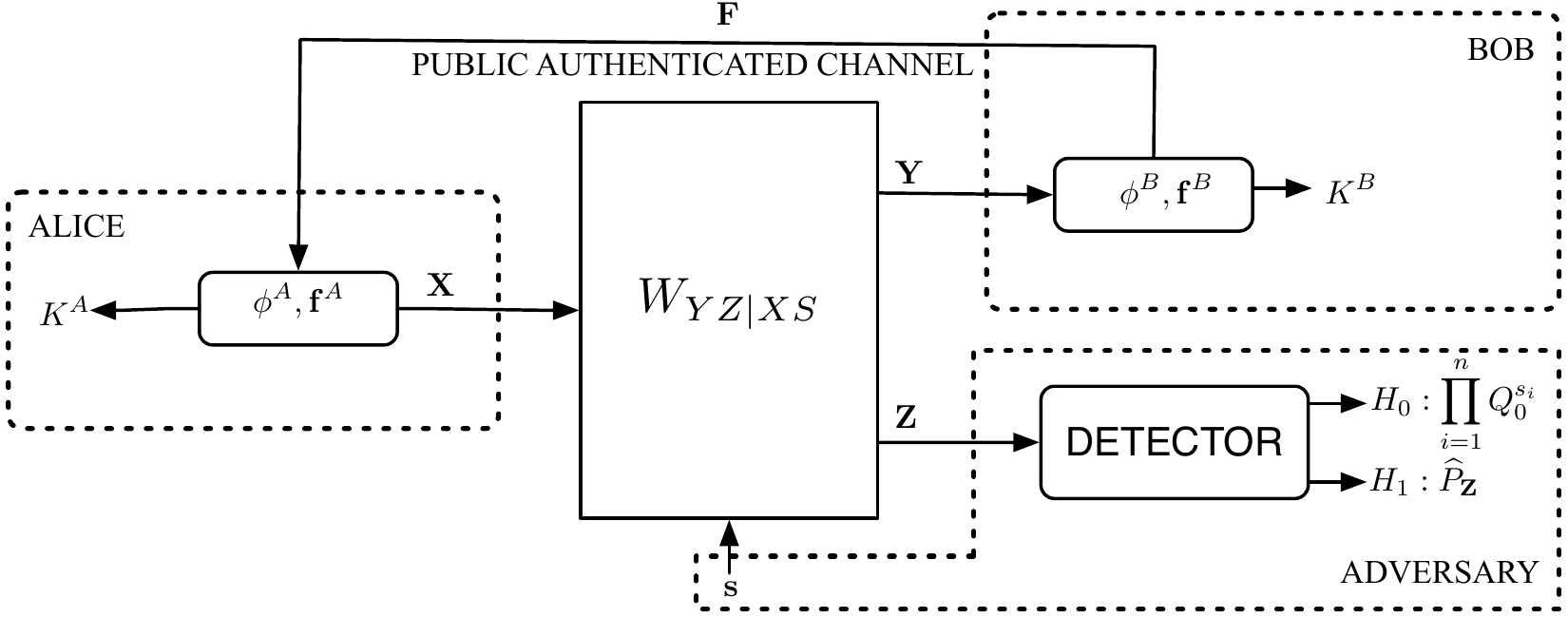}
  \caption{Covert secret key generation model}
  \label{fig:covert-skg-model}
\end{figure}

We consider the channel model for secret key generation illustrated in Fig.~\ref{fig:covert-skg-model}, in which two legitimate parties, Alice and Bob, attempt to generate a secret key while keeping the entire key generation process undetectable by a warden Willie. The channel is a state-dependent \ac{DMC} $(\calX\times\calS,W_{YZ|XS},\calY\times\calZ)$, in which the state $S$ is under Willie's control while the input $X$ is under Alice's control. Bob and Willie's channel outputs are $Y$ and $Z$, respectively. For simplicity, we assume that $\calX=\calS\eqdef\{0,1\}$, where $0$ is the input corresponding to the absence of communication. For $x,s\in\{0,1\}$, we define
\begin{align}
  P_x^s &\eqdef W_{Y|X=x,S=s},\quad Q^s_x \eqdef W_{Z|X=x,S=s},\displaybreak[0]\\
\text{ and }  (PQ)^s_x &\eqdef W_{YZ|X=x,S=s}.
\end{align}

Secret key generation is enabled by the presence of a public authenticated link of unlimited capacity, which Bob may use to transmit symbols in alphabet $\calF$. \modfirst{Alice and Bob have access to a source of secret common randomness $(\calR_C,P_{R_C})$ and each possess a local source of randomness denoted by $(\calR_A,P_{R_A})$, $(\calR_B,P_{R_B})$, respectively. To make the problem non-trivial, the entropy of $P_{R_C}$ is subtracted from the number of generated key bits in our throughput analysis.}

In the presence of an active attacker controlling the channel state, the final length of the key is not known ahead of time. Alice and Bob must therefore merely agree a priori on a maximum number of channel uses and  bits of secret key. Formally, a code for key generation $\calC$ over $n$ channel uses for a maximum of $m$ key bits consists of the following.
\begin{itemize}
\item $n$ encoding function for Alice $\mathbf{f}^A = (f_1^A,\cdots, f_n^A)$, where $f_i^A:\mathcal{F}^{i-1} \times \mathcal{R}^A\times \calR^C\to \mathcal{X}$ specifies the symbol sent by Alice at time $i\in\intseq{1}{n}$;
\item  $n$ encoding functions for Bob $\mathbf{f}^B = (f_1^B,\cdots, f_n^B)$, where $f_i^B:\mathcal{Y}^{i} \times \mathcal{R}^B\times \calR^C\to \modfirst{ \mathcal{F}}$ specifies the symbol sent by Bob at time $i\in\intseq{1}{n}$ over the public channel;
\item a key extraction function $\phi^A:\mathcal{F}^n \times \mathcal{R}^A \times \calR^C\to \{0, 1\}^m$ for Alice;
\item a key extraction function  $\phi^B:\mathcal{Y}^n \times \mathcal{R}^B \times \calR^C\to \{0, 1\}^m$ for Bob;
\item an estimator for the number of key bits $\ell^A:\mathcal{F}^n \times \mathcal{R}^A \times \calR^C\to \intseq{0}{m}$ for Alice;
\item an estimator for the number of key bits $\ell^B:\mathcal{Y}^n \times \mathcal{R}^B  \times \calR^C\to \intseq{0}{m}$ for Bob.
\end{itemize}
We assume that the protocol is known to all parties. The sequence of $n$ random symbols transmitted by Alice is denoted $\mathbf{X}\in\calX^n$, while the sequence of states is denoted $\mathbf{s}\in\calS^n$. Note that we do not require the existence of a probability distribution for the state sequence. The sequence of observations at Bob and Willie are denoted $\mathbf{Y}\in\calY^n$ and $\mathbf{Z}\in\calZ^n$, respectively. Bob's public communication is collectively denoted by $\mathbf{F}$ and the generated keys are denoted $K_A$, $K_B$, respectively. For a fixed state sequence $\mathbf{s}$, the distribution induced by the coding scheme is denoted $\widehat{P}_{\mathbf{X}\mathbf{Y}\mathbf{Z}\mathbf{F}K^AK^B\ell^A\ell^B|\mathbf{s}}$. \modfirst{Note that our model allows for an arbitrary $\mathbf{s}$ but does not allow the warden to adapt $\mathbf{s}$ to its observations.}

The performance of the key generation scheme is measured in terms of the following metrics:
\begin{itemize}
\item the probability of error $P_e(\mathcal{C}|\mathbf{s})\eqdef\P[\widehat{P}_{K^AK^B}]{\ell^A\neq \ell^B \text{ or } K^A(i) \neq K^B(i) \text{ for } i\in\intseq{1}{\ell^B}\big|\mathbf{s}}$, where $K^A(i)$ and $K^B(i)$ denote the $i^{th}$ bit in  Alice's key and Bob's key, respectively;
\item the secrecy $\D{\widehat{P}_{K^B\mathbf{F}\mathbf{Z}|\mathbf{s}}}{P_{K^B}^{\mathrm{unif}}\times \widehat{P}_{\mathbf{F}\mathbf{Z|\mathbf{s}}} }$;
\item the covertness $C(\mathcal{C}|\mathbf{s}) \eqdef \D{\widehat{P}_{\mathbf{F}\mathbf{Z}|\mathbf{s}}}{P_{\mathbf{F}}^{\mathrm{unif}} \times \prod_{i=1}^nQ_0^{s_i}}$.
\end{itemize}
We call $\mathcal{C}$ a $(2^m, n, r, \epsilon, \delta, \tau, \mathbf{s})$ code if $\avgH{R^C}\leq r$, $P_e(\mathcal{C}|\mathbf{s}) \leq \epsilon$, $S(\mathcal{C}|\mathbf{s}) \leq \delta$, and $C(\mathcal{C}|\mathbf{s})\leq \tau$. 

While the definitions of the probability of error and secrecy metrics are standard, there are somewhat arbitrary choices in our definition of covertness. We require Alice's transmission to be indistinguishable from an all-$0$ transmission, but we allow Bob to send symbols on the public channel as long as their observation does not help Willie's detection. \modfirst{For small $\tau>0$, the covertness constraint ensures that the public communication is nearly uniformly distributed and independent of the Willie's observation on the noisy channel.} One can think of Bob as a terminal emitting seemingly random ``beacons'' that do not divulge the existence of a secret key generation protocol.

The throughput achieved by a protocol is defined as follows.
\begin{definition}
A throughput $R$ is achievable \ac{wrt} the sequence $\{\mathbf{s}_n\}_{n\geq 1}$, if there exists a sequence of $(2^{m_n}, n, r_n, \epsilon_n, \delta_n, \tau_n, 
\mathbf{s}_n)$ codes $\{\mathcal{C}_n\}_{n\geq 1}$ such that
\begin{align}
\lim_{n\to \infty} \epsilon_n = \lim_{n\to\infty}\delta_n = \lim_{n\to\infty} \tau_n  = 0, ~m_n = \omega(\log n)
\end{align}
\begin{align}
\lim_{n\to \infty} \P{\frac{\ell^B_n-r_n}{\sqrt{n\tau_n}} \geq R} = 1.
\end{align}
\end{definition}

The special case of a \emph{passive attacker} consists of the situation in which the state sequence is fixed and known ahead of time. We set this sequence to be $\mathbf{s}=\mathbf{0}$ and drop the indices referring to the state to simplify notation. In this case, note that the estimators $\ell^A$ and $\ell^B$ are not needed and that the total number of key bits $m$ may be fixed ahead of time. We can then formally define the covert secret key capacity as follows.
\begin{definition}
\label{def:passive-code}
A throughput $R$ is achievable with a passive attacker if there exists a sequence of $(2^{m_n}, n, 0, \epsilon_n, \delta_n, \tau_n)$ codes $\{\mathcal{C}_n\}_{n\geq 1}$ such that
\begin{align}
\lim_{n\to \infty} \epsilon_n = \lim_{n\to\infty}\delta_n = \lim_{n\to\infty} \tau_n  = 0, ~m_n = \omega(\log n),
\end{align}
\begin{align}
\liminf_{n\to \infty} \frac{m_n}{\sqrt{n\tau_n}} \geq R.
\end{align}
The supremum of all achievable throughputs is denoted $C_{\text{csk}}$. 
\end{definition}
\modfirst{
\begin{remark}
\label{rem:arbitrary-alphabet}
Our restriction to $\card{\calS} = \card{\calX} = 2$ simplifies the technical details in our proofs. By following~\cite[Section VII-B]{Bloch2016a}, one can extend the results to any finite $\calX$. When $\card{S}>2$, one can adapt our estimation protocol to operate on the type of the state sequences instead of their weight.
\end{remark}
}

\section{Covert secret key capacity with a passive warden}
\label{sec:main-results-passive}
For completeness, we recall without proof the partial characterization of the covert secret key capacity in the presence of a passive warden, which is our main result from \cite{Tahmasbi2017}.
\begin{theorem}
\label{th:passive_general_bound}
If $(PQ)_0 = P_0 \times Q_0$, then
\begin{multline}
\label{eq:passive_achievability_bound}
\sqrt{\frac{2}{\chi_2(Q_1\|Q_0)}}\left(\D{(PQ)_1}{(PQ)_0} - \D{Q_1}{Q_0}\right) \geq   C_{\textnormal{csk}}\\
 \geq  \sqrt{\frac{2}{\chi_2(Q_1\|Q_0)}}\left(\D{(PQ)_1}{(PQ)_0}  - \D{Q_1}{Q_0}\right.\\
 \\ \left. - \D{(PQ)_1}{P_1\times Q_1}\right).
\end{multline}
\end{theorem}
\begin{IEEEproof}
See~\cite{Tahmasbi2017}.
\end{IEEEproof}
As an application of the above result, we characterize the exact covert secret key capacity when the channels from Alice to Bob and Willie are independent.
\begin{corollary}
  \label{cor:covert-secret-key}
  If $(PQ)_1 = P_1\times Q_1$ and $(PQ)_0 = P_0\times Q_0$,  then 
  \begin{align}
    C_{\textnormal{csk}} = \sqrt{\frac{2}{\chi_2(Q_1\|Q_0)}} \D{P_1}{P_0}.
  \end{align}
\end{corollary}
Corollary~\ref{cor:covert-secret-key} may be somewhat surprising in that it suggests that secrecy comes ``for free'' since the covert secret-key capacity is equal to the covert capacity of the channel. In practice, however, some small amount of privacy amplification would still be needed and the effect of the warden's channel only disappears in the asymptotic limit of large sequences. This result is an artifact of the model, which ensures that the information leakage from Bob to Willie has \emph{negligible} scaling compared to the information transfer from Bob to Alice.

\section{Covert throughput with an Active Warden}
\label{sec:main-results-active}
We now develop results for an active warden and show the existence of a sequence of coding schemes generating a key for \emph{any} sequence of states. The number of generated key bits depends on the state sequences only through their weights. 

\begin{theorem}
\label{th:main_active}
Let $(\calX\times \calS, W_{YZ|XS}, \calY, \calZ)$ be an arbitrarily varying \ac{DMC}  with $\calX = \calS = \{0, 1\}$, $P_0^0 = P_0^1 = P_0$, \modfirst{$P_1^1 \neq P_1^0$}, and $(PQ)_0^s = P_0^s\times Q_0^s$ for $s\in\calS$. There exists a sequence of codes $\{\calC_n\}_{n\geq 1}$ such that \modfirst{for all $\beta\in [0,1]$ }and all sequences $\{\mathbf{s}_n\}_{n\geq 1}$ with $\lim_{n\to\infty}\frac{\modfirst{\wt{\mathbf{s}_n}}}{n} = \beta$, the following covert throughput is achievable
\begin{align}
\label{eq:cov-active-rate}
R(\beta) \eqdef \sqrt{2}\frac{\D{(1-\beta)P_1^0+\beta P_1^1}{P_0} - \pr{(1-\beta)I^0+\beta I^1}}{\sqrt{(1-\beta) \chi_2(Q_1^0\|Q_0^0)+ \beta \chi_2(Q_1^1\|Q_0^1)}},
\end{align}
where  for $s\in\calS$,
\begin{multline}
I^s \eqdef \D{Q_1^s}{Q_0^s}+\D{P_1^s}{P_0^s}-\D{(PQ)_1^s}{(PQ)_0^s}\\
\modfirst{+}\D{(PQ)_1^s}{P_1^s\times Q_1^s}.
\end{multline}

\end{theorem}
\begin{remark}Comparing the throughputs in \eqref{eq:passive_achievability_bound} and \eqref{eq:cov-active-rate}, note that the quantities corresponding to the main channel in \eqref{eq:passive_achievability_bound} are replaced by the same quantities for the channel $\sum_{i=1}^n \frac{1}{n} W_{Y|XS=s_i}$, and the quantities corresponding to the warden's channel are replaced by the average of those quantities over the different channel uses. The intuition is that Alice and Bob have no direct access to the state sequences and only approximate the weight through their noisy observations. From their perspective, the best approximation of the main channel is $\sum_{i=1}^n \frac{1}{n} W_{Y|XS=s_i}$. In contrast, Willie knows the exact state sequence therefore obtains information from each realized channel use.
\end{remark}
\modfirst{
\begin{remark}
For an \ac{AVC}, one can eliminate the common randomness when the channel is non-symmetrizable as the capacity without common randomness is non-zero \cite{csiszar2011information}. However, such consideration does not directly apply to our model, which includes a feedback link from the receiver.
\end{remark}
We also establish a straightforward converse result for Theorem~\ref{th:main_active}. Note that it is challenging to obtain a converse that matches our achievability result for a couple of reasons. First, even without covertness constraint and active adversaries, it is generally hard to provide a tight converse for the secret key generation problem because of the interactions allowed by the existence of the public communication. Second, our achievability result proves the possibility of covert secret key generation for all channel state sequences, which makes the problem of finding a tight converse more difficult.
\begin{theorem}
\label{th:converse-active}
For any  sequence of codes $\{\calC_n\}_{n\geq 1}$ that achieves covert throughput $R$ over the channel in Theorem~\ref{th:main_active} for a state sequence $\{\mathbf{s}_n\}_{n\geq 1}$ with $\lim_{n\to  \infty} \frac{\wt{\mathbf{s}_n}}{n} = \beta$, we have
\begin{align}
R \leq \sqrt{2}\sqrt{\beta \frac{( \D{P_1^1}{P_0} - I^1)^2}{\chi_2(Q_1^0\|Q_0^0)} + (1-\beta)\frac{(\D{P_1^0}{P_0} - I^0)^2}{ \chi_2(Q_1^1\|Q_0^1)}},
\end{align}
where $I^0$ and $I^1$ are defined in Theorem~\ref{th:main_active}.
\end{theorem}
\begin{IEEEproof}
To obtain a converse, we can assume that Alice and Bob know the state sequence $\mathbf{s}_n = (s_1, \cdots, s_n)$ so that the result follows from the same argument as for the passive warden \cite{Tahmasbi2017}. In particular, let $\calC_n$ be a $(2^m, n, r, \epsilon, \delta, \tau, \mathbf{s})$. By \cite{Ahlswede1993}, we obtain that
\begin{align}
\sqrt{n\tau_n}R \leq (1+o(1)) \sum_{i=1}^n \avgI{X_i;Y_i|Z_i S_i = s_i}.
\end{align}
Let $J$ be uniformly distributed over $\intseq{1}{n}$ and $\mu^s =  \P{X_J = 1|S_J = s}$. Using \cite[Eq. (60) and Eq. (61)]{Tahmasbi2017} and $\wt{\mathbf{s}_n} = n(\beta + o(1))$, we obtain

\begin{multline}
R \leq (1+o(1)) \\
\times \frac{\beta \mu^1 \pr{ \D{P_1^1}{P_0} - I^1} + (1-\beta) \mu^0 \pr{ \D{P_1^0}{P_0} - I^0}}{\sqrt{\beta(\mu^1)^2 \chi_2(Q_1^1\| Q_0^1) + (1-\beta)(\mu^0)^2 \chi_2(Q_1^0\|Q_0^0)}}.
\end{multline}
Maximizing over $\mu^1$ and $\mu^0$ yields the desired result.

\end{IEEEproof}
\begin{figure}[h]
  \centering
  \includegraphics[width=\linewidth]{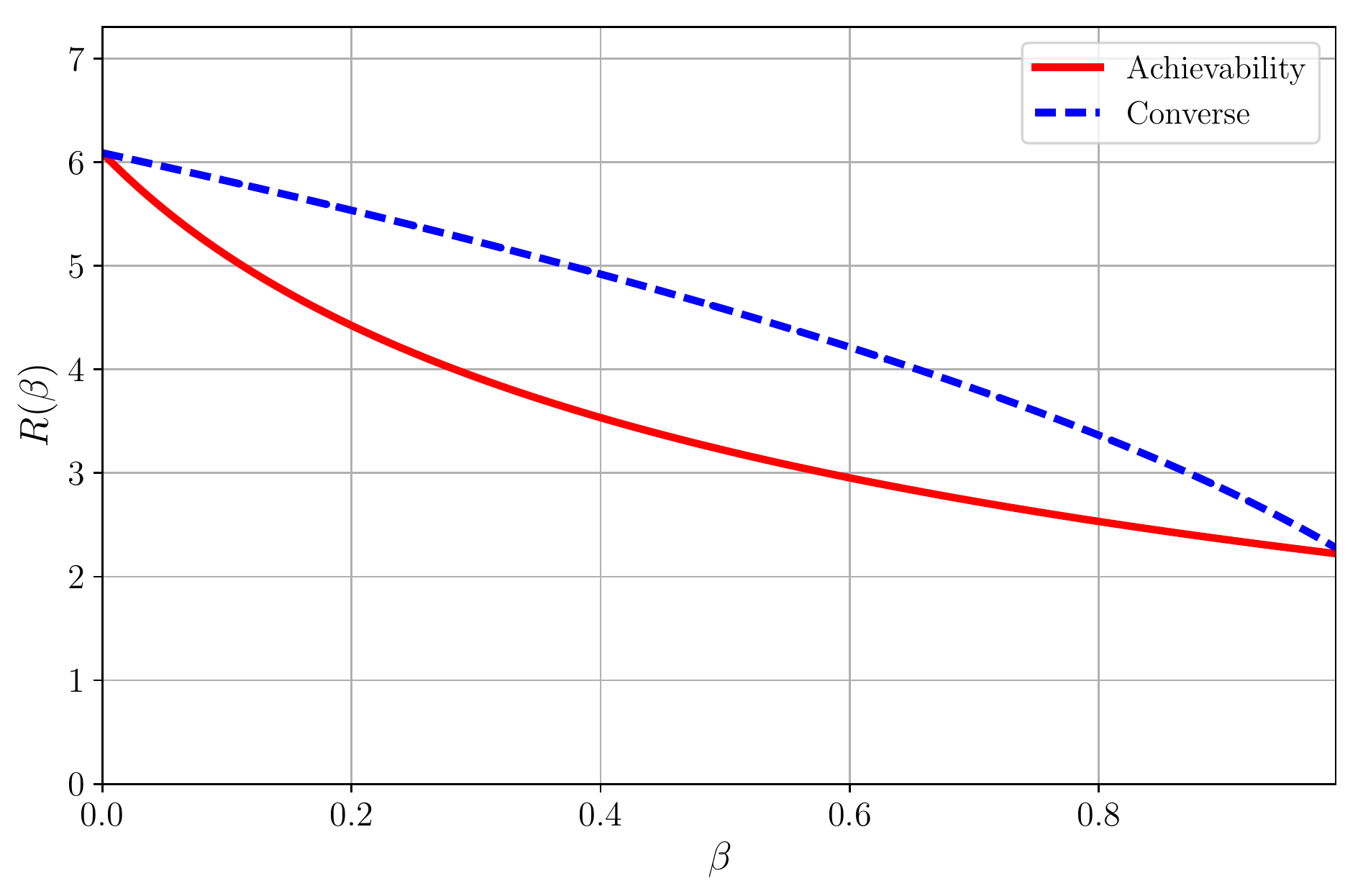}
  \caption{Illustration of Theorem~\ref{th:main_active} and Theorem~\ref{th:converse-active} for a \acp{BSC}}
  \label{fig:example}
\end{figure}

In Fig.~\ref{fig:example}, we illustrate the result of Theorem~\ref{th:main_active} and \ref{th:converse-active} for the
following example. When $s= 0$, Bob and Willie's channel are independent \ac{BSC}(0.1) and \ac{BSC}(0.4), respectively. When $s=1$, Bob's channel is a binary asymmetric channel with flipping probability 0.1 and 0.2 for $x=0$ and $x=1$, respectively, while Willie's channel is \ac{BSC}(0.3). The bounds are reasonably tight but, as expected, do not match.
}
\subsection{Proof of Theorem~\ref{th:main_active}}
We break down the proof of Theorem~\ref{th:main_active} into six steps.
\begin{enumerate}
\item We first establish a technical lemma pertaining to a concentration inequality for the reciprocal of the sum of \ac{iid} random variables.
\item We define an auxiliary problem and derive one-shot reliability and secrecy results.
\item We specialize the auxiliary problem to an arbitrarily varying \ac{DMC} and assume that an oracle provides the weight of the warden's state sequence. We use the result of Step 2 to develop a universal secrecy and reliability scheme. 
\item We reduce the amount of common randomness required for the coding scheme developed in Step 3.
\item We remove the oracle from \modfirst{the coding scheme} by introducing estimators for the weight of the warden's state sequence.
\item Finally, we combine all steps to prove the result.
\end{enumerate}

\subsubsection{A Concentration Inequality}
Suppose $\{X_i\}_{i=1}^n$ are \ac{iid} according to Bernoulli($p$). Since $\E{\sum_{i=1}^n X_i} = np$, one could expect  $\E{\left|\frac{1}{1+\sum_{i=1}^nX_i} - \frac{1}{(n+1)p}\right| }$ to be small. The following lemma formalizes this intuition and \modfirst{is proved in Appendix~\ref{sec:proof-bernoulli_con}.}
\begin{lemma}
\label{lm:bernoulli_con}
Suppose $X_1, \cdots, X_n$ are \ac{iid} according to Bernoulli($p$). Then, for $\frac{2}{np}<\epsilon < 1$,
\begin{multline}
\label{eq:concen_p}
\P{\left|\frac{1}{1+\sum_{i=1}^nX_i } - \frac{1}{(n+1)p}\right| \geq \frac{\epsilon}{(n+1)p}}\\
 \leq 2\exp\pr{-\frac{np\epsilon^2}{32}},
\end{multline}
\begin{multline}
\label{eq:concen_e}
\E{\left|\frac{1}{1+\sum_{i=1}^nX_i} - \frac{1}{(n+1)p}\right| }\\
 \leq \frac{\epsilon}{(n+1)p} + \left(1 + \frac{1}{(n+1)p}\right)e^{-\frac{np\epsilon^2}{32}}.
\end{multline}
\end{lemma}

\subsubsection{One-shot Results for an Auxiliary Problem}
\label{sec:oneshot-aux}
We introduce an auxiliary problem with the help of which we solve the main problem later on. \modfirst{The main rationale for introducing the auxiliary problem is to use a variation of the likelihood encoder~\cite{Song2016}, which allows us to exploit channel coding tools. This helps us avoid a finite length penalty that would appear if using source coding tools and would dominate the covert throughput.} Alice, Bob and Willie have access to $X\in\calX$, $Y\in\calY$ and $Z\in\calZ$, respectively, with joint distribution $P_{XYZ}$. 
\modfirst{In addition, Alice and Bob share secret common randomness in the form of a random codebook $\widetilde{\mathbf{Y}} =\{\widetilde{Y}_{w_1w_2}\}_{w_1\in\intseq{1}{M_1}, w_2\in\intseq{1}{M_2}}\in\calY^{M_1M_2}$ distributed according to $Q_Y^{\proddist M_1M_2}$ for some chosen $Q_Y$. Bob generates two messages $W_1\in\intseq{1}{M_1}$ and $W_2\in\intseq{1}{M_2}$ from $Y$ and  $\widetilde{\mathbf{Y}}$ according to the conditional \ac{PMF}
\begin{multline}
P_{W_1W_2|Y\widetilde{\mathbf{Y}}}(w_1, w_2|y, \widetilde{\mathbf{y}}) \\=  \begin{cases} \frac{\indic{y=\widetilde{y}_{w_1w_2}}}{\sum_{w_1'w_2'} \indic{y=\widetilde{y}_{w_1'w_2'}}}\quad &\sum_{w_1'w_2'} \indic{y=\widetilde{y}_{w_1'w_2'}} \neq 0 \\
\frac{1}{M_1M_2}\quad &
\sum_{w_1'w_2'} \indic{y=\widetilde{y}_{w_1'w_2'}} = 0
 \end{cases}.
\end{multline}
This operation induces the joint distribution $P_{XYZ\modfirst{\widetilde{\mathbf{Y}}}W_1W_2} \eqdef P_{XYZ}Q_{Y}^{\proddist M_1M_2}P_{W_1W_2|Y\mathbf{\widetilde{Y}}}$. Note that $Q_Y$ may be different from $P_Y$, which will be useful when analyzing universality and constitutes the main deviation from the standard likelihood encoder. The next two lemmas provide bounds showing that $W_1$ and $W_2$ can be interpreted as a secret key and a public message, respectively.} 
\begin{lemma}
\label{lm:oneshot_active_rel}
\modfirst{Let $\nu:\calX \times \calY \to \mathbb{R}$ be a fixed function and define a \emph{universal} decoder for estimating $W_1$ from $X, \widetilde{\mathbf{Y}}, W_2$ as 
\begin{align}
\hat{w}_1= \phi(x, \widetilde{\mathbf{y}}, w_2) \eqdef \mathop{\arg \max}_{w_1} \nu \pr{x, \widetilde{y}_{w_1w_2}}
\end{align}}
For all $\gamma > 0$, $\mu_Q\eqdef \min_yQ_Y(y)$, and $\frac{2}{(M_1M_2-1)\mu_Q}<\delta<1$, we have 
\begin{multline}
\P{W_1 \neq \widehat{W}_1} \leq \sum_{x, y}P_{XY}(x, y)\min(1, M_1q(x, y)) \\
+ (2 + M_1M_2)e^{-\frac{(M_1M_2-1)\mu_Q\delta^2}{32}}+\delta  ,
\end{multline}
where $q(x, y) \eqdef \sum_{y'}Q_Y(y') \indic{\nu(x, y')\geq \nu(x, y)}$. 
\end{lemma}
\begin{IEEEproof}
  See Appendix~\ref{sec:proof-oneshot_active_rel}.
\end{IEEEproof}
\begin{lemma}
\label{lm:oneshot_active_sec}
For all $\gamma > 0$ and $\frac{2}{(M_1M_2-1)\mu_Q}<\delta<1$, we have
\begin{multline}
\V{{P}_{W_1Z\widetilde{\mathbf{Y}}}, P_{W_1}^{\mathrm{unif}}\times{P}_{Z\widetilde{\mathbf{Y}}} } \leq \\\sum_{y, z} P_{YZ}(y, z) \indic{P_{YZ}(y, z) \geq \gamma P_Z(z) Q_Y(y)} +\frac{1}{2}\sqrt{\frac{\gamma}{M_2}} \\
+ \frac{1}{2}\delta + \frac{1}{2}(2 + M_1M_2)e^{-\frac{(M_1M_2-1)\mu_Q\delta^2}{32}}.
\end{multline}
\end{lemma}
\begin{IEEEproof}
  See Appendix~\ref{sec:proof-oneshot_active_sec}.
\end{IEEEproof}

\subsubsection{Universal Asymptotic Results for the Auxiliary Problem}
\label{sec:asymptotic-aux}

\modfirst{We now extend the auxiliary problem results of Section~\ref{sec:oneshot-aux} by using the channel $n$ times  allowing the warden to vary the channel at every channel use.}  More precisely, we consider an arbitrarily varying \ac{DMC} $(\calX\times \calS, W_{YZ|XS}, \calY, \calZ)$ with $\calX = \calS = \{0, 1\}$ and $P_0^0 = P_0^1 = P_0$.  For simplicity, we suppose for now that the weight $\wt{\mathbf{s}}$ of Willie's state sequence is provided to Alice and Bob by some \emph{oracle} at the end of the transmission. 
Alice samples the input sequence $\mathbf{X}$  according to $Q_X^\pn$ where $Q_X =\text{Bernoulli}(\alpha_n)$ and $\alpha_n \in \omega\left(\frac{\log n}{{n}}\right)\cap o\pr{\frac{1}{\sqrt{n}}}$\modfirst{, and transmits it over the channel so that Bob and Willie observe $\mathbf{Y}$ and $\mathbf{Z}$, respectively. Alice and Bob are assumed to share a random codebook $\{\widetilde{\mathbf{Y}}_{w_1, w_2, w_3}\}_{w_1\in\intseq{1}{M_{1}},w_2\in\intseq{1}{M_{2}},w_3\in\intseq{1}{M_{3}}}$ distributed \ac{iid} according to $P_0^\pn$. Bob randomly creates an encoder $F$ to generate $W_1$ and $W_2$ from $\mathbf{y}$ using   $\{\tilde{\mathbf{Y}_{w_1w_2w_3}}\}$ as 
\begin{align}
  \P{F(\mathbf{y}) = (w_1, w_2)} =   \frac{\sum_{w_3}\indic{\mathbf{y} = \widetilde{\mathbf{Y}}_{w_1w_2\modfirst{w_3}}}}{\sum_{w_1'w_2'w_3'}\indic{\mathbf{y} = \widetilde{\mathbf{Y}}_{w_1'w_2'w_3'}}}
\end{align}
if $\sum_{w_1'w_2'w_3'}\indic{\mathbf{y} = \widetilde{\mathbf{Y}}_{w_1'w_2'w_3'}} \neq 0$ and $\frac{1}{M_1M_2}$ else. Message $W_2$ is sent over the public channel and Alice subsequently uses $\mathbf{X}$ and $W_2$ to decode $W_1$ as $\widehat{W}_1$ with the random decoder $\Phi:\calX^n \times \intseq{1}{M_2} \to \intseq{1}{M_1}$ defined as 
\begin{align}
\Phi(\mathbf{x}, w_2) \eqdef \mathop{\arg\max}_{w_1}\pr{\max_{w_3} I(\mathbf{x}\wedge \widetilde{\mathbf{Y}}_{w_1w_2w_3})}.
\end{align}
}
For a code $(f, \phi)$ and a sequence of states $\mathbf{s}$, we define
\begin{align}
P_e(f, \phi | \mathbf{s}) &\eqdef \P{W_1\neq \widehat{W}_1|\mathbf{s}},\\
S(f, \phi |\mathbf{s}) &\eqdef\V{ \widehat{P}_{W_1W_2\mathbf{Z}|\mathbf{s}}, \widehat{P}_{W_1W_2}\times \widehat{P}_{\mathbf{Z}|\mathbf{s}}}.
\end{align}

 \modfirst{We shall now prove that the above random code performs well over a class of state sequences specified as follows. For each $\mathbf{s}$, set $\beta\eqdef \frac{\wt{\mathbf{s}}}{n}$, $Q_S \eqdef \text{Bernoulli}(\beta)$, and define the \ac{PMF}
\begin{align}
\label{eq:def-qsxyz}
Q^{\beta}_{SXYZ}(s,x, y, z) \eqdef Q_S(s)Q_X(x)W_{YZ|XS}(y,z|x,s).
\end{align}
Define $\calS(M_{1}, M_{2}, M_{3})$ as the set of state sequences $\mathbf{s}$ such that the corresponding $Q^{\beta}_{SXYZ}(s,x, y, z)$ satisfies}
\begin{align}
  \label{eq:M_123_bound}\log M_{1} + \log M_{2} + \log M_{3} &\geq \lceil (1+\zeta) \log \frac{1}{\mu_0}n\rceil,\\
  \label{eq:M_13_bound}\log M_{1} + \log M_{3} &\leq \lfloor(1 - \zeta) \avgI{X;Y} n\rfloor,\\
  \label{eq:M_3_bound}\log M_{3} \geq \lceil (1+\zeta) & \alpha_n\pr{\beta I^1 + (1-\beta)I^0} n \rceil.
\end{align} 
\modfirst{Note that $W_1$ should be interpreted as a secret key and $W_2$ should be interpreted as a public message. $W_3$ should be interpreted as additional randomization, which plays no role beyond helping us control the specific distribution used by the likelihood encoder. The next lemma shows that the random code described above is universal over the set $\calS(M_{1}, M_{2}, M_{3})$. Specifically, we prove that given the choice of $(M_1, M_2, M_3)$, there exists a protocol that performs well for all state sequences $\mathbf{s} \in \calS(M_{1}, M_{2}, M_{3})$. In Section~\ref{sec:estimation}, where we choose the values of $M_1$, $M_2$, and $M_3$ based on Bob's observations, we show that with high probability the true state sequence is in $\calS(M_{1}, M_{2}, M_{3})$.}
\begin{lemma}
\label{lm:active_random_code}
For all $\beta,\zeta>0$, there exists $\xi>0$ such that for large enough $n$ and for all $\mathbf{s} \in \calS(M_{1}, M_{2}, M_{3})$, we have
\begin{align}
\E[F, \Phi]{P_e(F\modfirst{,} \Phi |\mathbf{s})} &\leq 2^{-\omega(\log n)} \text{ and }\\
\E[F, \Phi]{S(F, \Phi |\mathbf{s})} &\leq 2^{-\xi \alpha_n n},
\end{align}
where the term $\omega(\log n)$ depends on $\zeta$ and the channel.
\end{lemma}
\begin{IEEEproof}
  See Appendix~\ref{sec:proof-active_random_code}.
\end{IEEEproof}

\subsubsection{Common Randomness Reduction}
In the next lemma, \modfirst{we use Ahlswede's \modfirst{elimination} technique \cite{Ahlswede1978} to reduce the amount of common randomness in the coding  scheme.}
\begin{lemma}
\label{lm:random-red}
Let $(F, \Phi)$ be any random code and $\calS_n$ be a subset of $\calS^n$. Furthermore, for all $\mathbf{s}\in \calS_n$, assume that $\E[F, \Phi]{P_e(F, \Phi|\mathbf{s})} \leq \epsilon$ and $\E[F, \Phi]{S(F, \Phi|\mathbf{s})} \leq \epsilon$. Then, there exist $L$ realizations $(f_1, \phi_1), \cdots, (f_L, \phi_L)$of the random code that satisfy
\begin{align}
\frac{1}{L}\sum_{i=1}^L P_e(f_i, \phi_i|\mathbf{s}) \leq \epsilon' \text { and }
\frac{1}{L}\sum_{i=1}^L S(f_i, \phi_i|\mathbf{s}) \leq \epsilon'
\end{align}
for all $\mathbf{s} \in \calS_n$ provided that
\begin{align}
\epsilon' > 2 \log(1 + \epsilon) \text{ and } L > \frac{2}{\epsilon'}(1 + n). \label{eq:robost_cond}
\end{align}
\end{lemma}
\begin{IEEEproof}
Let $(F_1, \Phi_1), \cdots, (F_L, \Phi_L)$ be $L$ \ac{iid} random codes distributed according to $P_{(F, \Phi)}$. For any $\mathbf{s}\in \calS_n$, we have
\begin{align}
&\P[F, \Phi]{\frac{1}{L}\sum_{i=1}^L P_e(F_i, \Phi_i|\mathbf{s}) \geq \epsilon' \text { or }\frac{1}{L}\sum_{i=1}^L S(F_i, \Phi_i|\mathbf{s}) \geq \epsilon'} \\\displaybreak[0]
&\leq 2^{-L\epsilon'}\E{2^{\sum_{i=1}^L P_e(F_i, \Phi_i|\mathbf{s}) }}+2^{-L\epsilon'}\E{2^{\sum_{i=1}^L S(F_i, \Phi_i|\mathbf{s}) }}\\\displaybreak[0]
&= 2^{-L\epsilon'}\pr{\E{2^{ P_e(F, \Phi|\mathbf{s}) }}}^L+2^{-L\epsilon'}\pr{\E{2^{ S(F, \Phi|\mathbf{s}) }}}^L\\
&\stackrel{(a)}{\leq} 2^{-L\epsilon'}\pr{1+\E{ P_e(F, \Phi|\mathbf{s}) }}^L+2^{-L\epsilon'}\pr{1+\E{ S(F, \Phi|\mathbf{s}) }}^L\\\displaybreak[0]
&\leq 2^{-L\epsilon'}\pr{1+\epsilon}^L+2^{-L\epsilon'}\pr{1+\epsilon}^L\\\displaybreak[0]
&= 2^{-L(\epsilon' - \log(1+\epsilon)) + 1}\modfirst{,}
\end{align}
\modfirst{where $(a)$ follows from $2^x \leq 1 + x$ for $x\in[0, 1]$.}
Therefore, the union bound yields that
\begin{multline*}
\mathbb{P}_{F, \Phi}\left(\forall  \mathbf{s}\in\calS_n, \frac{1}{L}\sum_{i=1}^L P_e(F_i, \Phi_i|\mathbf{s}) < \epsilon' \right.\\
\left.\text { and }\frac{1}{L}\sum_{i=1}^L S(F_i, \Phi_i|\mathbf{s}) < \epsilon'\right) > 1 - 2^n 2^{-L(\epsilon' - \log(1+\epsilon)) + 1}
\end{multline*}
which is positive given that \eqref{eq:robost_cond} holds.
\end{IEEEproof}

\begin{corollary}
\label{cor:random-red}
Under the same assumptions as Lemma~\ref{lm:active_random_code}, for all $\zeta>0$, all large enough $n$, and $L>  2n^4(1+n)$, there \modfirst{exist} codes $(f_1, \phi_1), \cdots, (f_L, \phi_L)$ such that for any $\mathbf{s}\in \calS_\beta(M_1, M_2, M_3)$,
\begin{align}
\frac{1}{L}\sum_{\ell = 1}^LP_e(f_\ell, \phi_\ell|\mathbf{s}) \leq \frac{1}{n^4},\text{ and }
\frac{1}{L}\sum_{\ell = 1}^LS(f_\ell, \phi_\ell|\mathbf{s}) \leq \frac{1}{n^4}. \label{eq:red-rand-perf}
\end{align}
\end{corollary}

\begin{IEEEproof}
We first consider the random code $(F, \Phi)$ introduced in Section~\ref{sec:asymptotic-aux} for which  we have $\E[F, \Phi]{P_e(F\modfirst{,} \Phi|\mathbf{s})} \leq 2^{-\omega(\log n)}$ and $\E[F, \Phi]{S(F, \Phi |\mathbf{s})} \leq 2^{-\xi \alpha_n n} = 2^{-\omega(\log n)}$ by Lemma~\ref{lm:active_random_code} for all $\mathbf{s}\in\calS(M_1, M_2, M_3)$. Applying Lemma~\ref{lm:random-red} to $(F, \Phi)$ for $L>  2n^4(1+n)$, we obtain $L$ codes $(f_1, \phi_1), \cdots, (f_L, \phi_L)$ such that \eqref{eq:red-rand-perf}  holds \modfirst{since the two constraints $n^{-4}\geq 2\log(1 + 2^{-\omega(\log n)})$ and \modfirst{$L > \frac{2}{n^{-4}}(1+n)$} in Lemma~\ref{lm:random-red} hold for large $n$.}
\end{IEEEproof}

\subsubsection{Estimation of the state sequence weigth}
\label{sec:estimation}
\modfirst{We now construct an estimator for $\wt{\mathbf{s}}$ to replace the oracle. This requires running the protocol over $n'= n + g$ channel uses, where $g$ is a positive integer to be specified later used to provision for channel estimation. Before transmission, Alice and Bob secretly and independently select every channel use for estimation with probability $\kappa_n\in[0, 1]$, which requires $n \Hb{\kappa_n}$ bits of shared secret key. 
Let $L$ denote the number of positions chosen for the estimation and let $\mathbf{J} = (J_1, \cdots, J_L)$ denote the corresponding indices in increasing order. If $n'-n\eqdef g < L$, Alice and Bob halt the protocol and do not generate a key. Otherwise, Alice transmits symbol ``1'' in the positions in $\mathbf{J}$ and operates as in Step 3 in the known $n$ positions not in $\mathbf{J}$. Since $P_0^1 \neq P_1^1$, there exists $y_0 \in \calY$ such that $P_0^1(y_0) \neq P_1^1(y_0)$. For $\mu_0 \eqdef P_0^1(y_0)$, $\mu_1 \eqdef P_1^1(y_0)$, and $T_i \eqdef \frac{\indic{Y_{j_i}=y_0} - \mu_0 }{\mu_1 - \mu_0 }$, Bob estimates $\beta = \frac{\wt{\mathbf{s}}}{n}$ as $\widehat{\beta} \eqdef \frac{1}{L} \sum_{i=1}^L T_i$ (for $L=0$, we define $\widehat{\beta} = 1$). \modfirst{Note that for a fixed $j_i$, $\E{T_i} = s_{j_i}$, and by the results on sampling without replacement, one can expect that $\sum_{i=1}^L T_i/L \approx \sum_{i=1}^L s_{J_i}/L  \approx \sum_{i=1}^n s_{i}/n =  \beta$.  }}

We now show that, with high probability, Alice and Bob do not halt the protocol and $\widehat{\beta}$ is close to $\beta$. With $g = (1+\mu) \kappa_n n'$, application of a Chernoff bound yields that $\P{L \geq g} \leq 2^{-\frac{\mu^2\kappa_nn'}{3}}$. In addition, for all $\lambda>0$ and $\mu\in]0,1[$, 
\begin{align}
 &\P{|\widehat{\beta} - \beta| > \lambda} \\\displaybreak[0]
 &= \sum_{\ell = 0}^{n'} \P{L = \ell}\P{\left|\widehat{\beta}- \beta \right| > \lambda \big | L = \ell }\\
 &\leq \P{L \leq (1-\mu)\kappa_nn'}\nonumber\\\displaybreak[0]
 & + \sum_{\ell = \lfloor(1-\mu)\kappa_nn'\rfloor + 1}^{n'}  \P{L = \ell}\P{\left|\frac{1}{\ell}\sum_{i=1}^\ell T_i - \beta \right| > \lambda \bigg | L = \ell }\\\displaybreak[0]
 &\stackrel{(a)}{\leq} \exp\pr{-\frac{1}{2}\mu^2 \kappa_n n'}\nonumber\\\displaybreak[0]
 &+ \sum_{\ell = \lfloor(1-\mu)\kappa_nn'\rfloor + 1}^{n'}  \P{L = \ell}\P{\left|\frac{1}{\ell}\sum_{i=1}^\ell T_i - \beta \right| > \lambda \bigg | L = \ell },\label{eq:condition-ell}
\end{align}
where $(a)$ follows from a Chernoff bound. Conditioned on $L = \ell$, $\mathbf{J} = (J_1, \cdots, J_\ell)$ is distributed uniformly on $\calJ^\ell = \{\mathbf{j} = (j_1, \cdots, j_\ell): j_1<\cdots<j_\ell\}$.  Upon defining the event
\begin{align}
\calE \eqdef \left\{\mathbf{j}\in\calJ^\ell:\left| \frac{1}{\ell}\sum_{i=1}^\ell s_{j_i}-\beta\right| > \frac{\lambda}{2}\right\},
\end{align}
we have
\begin{align}
&\P{\left|\frac{1}{\ell}\sum_{i=1}^\ell T_i - \beta \right| > \lambda \bigg | L = \ell}\displaybreak[0]\\
&= \P{\left|\frac{1}{\ell}\sum_{i=1}^\ell T_i - \beta \right| > \lambda \bigg | L = \ell, \calE}\P{\calE|L = \ell} \displaybreak[0]\\
& +  \P{\left|\frac{1}{\ell}\sum_{i=1}^\ell T_i - \beta \right| > \lambda \bigg | L = \ell, \calE^c}\P{\calE^c|L = \ell}\displaybreak[0]\\
&\leq \P{\calE|L = \ell} + \P{\left|\frac{1}{\ell}\sum_{i=1}^\ell T_i - \beta \right| > \lambda \bigg | L = \ell, \calE^c}.\displaybreak[0]
\end{align}
We next express $\P{\calE|L = \ell}$ as the \ac{CDF} of the hypergeometric distribution. In particular, let $H$ denote the number of successes in $\ell$ draws without replacement from a population of size $n'$ with $\wt{\mathbf{s}} = \ell\beta $ successes in the population. We then have 
\begin{align}
\P{\calE|L = \ell}=\P{\left|\frac{1}{\ell}H - \beta\right| \geq \frac{\lambda}{2}} \stackrel{(a)}{\leq} \exp\pr{-\frac{\lambda^2\ell}{2}},
\end{align}
where $(a)$ follows from the standard tail bounds for hypergeometric distribution (e.g., see \cite{hoeffding1963probability}).

We next fix some $\mathbf{j}\in \calJ^\ell \setminus \calE$. Since $\E{T_i|J_i = j_i} = s_{j_i}$, and $\frac{ - \mu_0}{\mu_1-\mu_0 } \leq T_i \leq \frac{1 - \mu_0 }{\mu_1 -\mu_0 }$, Hoeffding's inequality implies that
\begin{align}
&\P{|\widehat{\beta} - \beta| > \lambda |L = \ell, \mathbf{J} = \mathbf{j}}\\\displaybreak[0]
&= \P{\left|\frac{1}{\ell}\sum_{i=1}^\ell T_i - \beta \right|\geq \lambda  \bigg| L = \ell,\mathbf{J} = \mathbf{j}}\\\displaybreak[0]
&\leq 2\exp\pr{-2\ell(\mu_1-\mu_0)^2\pr{\lambda - \left|\beta - \frac{1}{\ell}\sum_{i=1}^\ell s_{j_i} \right|}^2}\\\displaybreak[0]
&\stackrel{(a)}{\leq}  2\exp\pr{-\frac{\ell(\mu_1-\mu_0)^2\lambda^2}{2}},
\end{align}
where $(a)$ follows from $\mathbf{j}\notin \calE$. Therefore, we obtain
\begin{multline}
 \P{|\widehat{\beta} - \beta| > \lambda|L=\ell}\\
 \leq  2\exp\pr{-\frac{(\mu_1-\mu_0)^2\lambda^2\ell}{2}} + 2 \exp\pr{-\frac{\lambda^2\ell}{2}}, \label{eq:bound_beta}
\end{multline}
which is less than $2^{-\xi \ell}$ for some $\xi > 0$ small enough and independent of $\ell$. \modfirst{Combining~\eqref{eq:bound_beta} with \eqref{eq:condition-ell},} we obtain $\P{|\widehat{\beta} - \beta| > \lambda} \leq 2^{-\xi \kappa_n n'}$ for some $\xi>0$ small enough.

We next show that for $Q_{\widehat{S}}=$ Bernoulli($\widehat{\beta}$), \modfirst{$Q^{\widehat{\beta}}_{\widehat{S}\widehat{X}\widehat{Y}\widehat{Z}}$} defined as \modfirst{in} \eqref{eq:def-qsxyz} 
and $(M_1, M_2, M_3)$ such that\modfirst{
\begin{align}
\label{eq:M_123}\log M_{1} + \log M_{2} + \log M_{3} &\geq \lceil (1+\zeta) \log \frac{1}{\mu_0}n\rceil,
\end{align}
\begin{align}
\label{eq:M_13}\log M_{1} + \log M_{3} &\leq \lfloor(1 - \zeta) \pr{\avgI{\smash{\widehat{X};\widehat{Y}}} - \zeta\alpha_n} n\rfloor,
\end{align}
\begin{align}
\label{eq:M_3}\log M_{3} &\geq \left\lceil(1+\zeta)\alpha_n\pr{ \widehat{\beta} {I}^1 + (1-\widehat{\beta}){I}^0 + \zeta }n \right \rceil,
\end{align}}
we have $\mathbf{s}\in \calS(M_1, M_2, M_3)$ with high probability. To do so, we should verify that \eqref{eq:M_123_bound}-\eqref{eq:M_3_bound} hold with high probability. By our definition of $(M_1, M_2, M_3)$, \eqref{eq:M_123_bound} is always true. Additionally, for $Q_{{S}}=$ Bernoulli(${\beta}$), and $Q^\beta_{SXYZ}$ defined as \modfirst{in} \eqref{eq:def-qsxyz}, the function
\begin{multline}
\psi(\beta)\eqdef\mathbb{D}(\beta W_{Y|X=1S=1}+(1-\beta)W_{Y|X=1S=0}\\ \|\beta W_{Y|X=0S=1}+(1-\beta)W_{Y|X=0S=0})
\end{multline}
 is continuous  in $\beta$. Therefore, there exists  $\lambda > 0$ such that if $|\beta - \beta'| < \lambda$,  $|\psi(\beta) - \psi(\beta')| \leq \zeta - o(1)$. Then,
\begin{multline}
\P{ \lfloor(1 - \zeta) \pr{\avgI{\smash{\widehat{X};\widehat{Y}}} - \zeta\alpha_n} n\rfloor \geq \lfloor(1 - \zeta) \avgI{{X};{Y}} n\rfloor  }\\
\begin{split}
&= \P{ {\avgI{\smash{\widehat{X};\widehat{Y}}} - \zeta\alpha_n} \geq  \avgI{{X};{Y}}  + O\pr{\frac{1}{n}}  }\\
&=  \P{ \alpha_n \psi(\widehat{\beta})- \zeta\alpha_n \geq  \alpha_n \psi(\beta)  + O\pr{\frac{1}{n}}  }\\
&\leq \P{|\psi(\beta) - \psi(\widehat{\beta})|\geq \zeta - o(1)}\\
&\leq \P{|\beta - \widehat{\beta}| \geq \lambda}\\
&\leq 2^{-\xi \kappa_nn'}.
\end{split}
\end{multline}
Similarly, we can argue that 
\begin{multline}
\mathbb{P}\left( \left\lceil(1+\zeta)\alpha_n\pr{ \widehat{\beta} {I}^1 + (1-\widehat{\beta}){I}^0 + \zeta }n \right \rceil\right. \\ \left.\leq  \left\lceil(1+\zeta)\alpha_n\pr{ \beta {I}^1 + (1-\beta){I}^0  }n \right \rceil \right)\leq 2^{-\xi \kappa_n n'},
\end{multline}
so that 
\begin{multline}
\mathbb{P}\left(\log M_1 \leq (1 - \zeta) \pr{\avgI{{X};{Y}} - \zeta\alpha_n} n- (1+\zeta) \right.\\
\left. \times \alpha_n\pr{ \beta {I}^1 + (1-\beta){I}^0 + \zeta }n - \zeta \alpha_n n\right) \leq 2^{-\xi \kappa_n n'}.\label{eq:logM1-bound}
\end{multline}
Hence, $\mathbf{s}\in \calS(M_1, M_2, M_3)$ with probability more than $1-2^{-\xi  \kappa_nn' + 1}$.

\subsubsection{Proof of Theorem~\ref{th:main_active}}
We put together the different pieces developed so far and describe  our active covert key generation protocol.  Let $\zeta > 0$, $n' = n + g$ be the block-length, $\kappa_n = o(\alpha_n/\log n) \cap \omega(\log n /n)$, $g\geq (1+\mu)\kappa_n n'$ for some $\mu \in ]0, 1[$, $\mathbf{J} = (J_1, \cdots, J_L)$ be the positions to be used for the estimation, and $K$ be a shared secret key uniformly distributed  over $\intseq{1}{U}$ for any $U> 2n^4(n+1)$. For $L > g$, the protocol halts. Otherwise, Alice samples $\mathbf{\widetilde{X}}$ according to $Q_X^\pn$ \modfirst{for transmission over} the channel $W_{YZ|XS}$ at the $n$ positions not included in $\mathbf{J}$, and transmits $1$ \modfirst{in} the positions in $\mathbf{J}$. Let $\mathbf{Y}$ and $\mathbf{Z}$ denote Bob's and Willie's received sequences, respectively, and $\widetilde{\mathbf{Y}}$ denote the sub-sequence of $\mathbf{Y}$  obtained by removing the \modfirst{components} in $\mathbf{J}$. Bob first estimates the type $\widehat{\beta}$ of Willie's states sequence defined in Section~\ref{sec:estimation} and sets $(M_1, M_2, M_3)$ such that \eqref{eq:M_123}-\eqref{eq:M_3} hold. Subsequently, for $(f_1, \phi_1), \cdots, (f_U, \phi_U)$ defined in Corollary~\ref{cor:random-red}, Bob generates two messages $(W_1, W_2) = f_{K}(\mathbf{\widetilde{Y}})$ and broadcasts $W_2$ together with \modfirst{$\widehat{\beta}$} one-time-padded with a shared secret key. Finally, Alice decodes $W_1$ as $\widehat{W}_1 \eqdef \phi_{K}(\mathbf{\widetilde{X}}, W_2)$. We provide the performance analysis of the protocol in four parts.

\paragraph{Reliability analysis} With probability at most $2^{-\xi \kappa_n n'} \leq 2^{-\omega(\log n)}$, the protocol is halted. If $\mathbf{s}\in \calS(M_1, M_2, M_3)$, then by Corollary~\ref{cor:random-red}, the probability of error is less than $\frac{1}{n^4}$. Since $\P{\mathbf{s}\in \calS(M_1, M_2, M_3)} \geq 1 - 2^{-\xi \kappa_n n' + 1} \leq 2^{-\omega(\log n)}$, the probability of error for the protocol is less than $n^{-4} + 2^{-\omega(\log n)}$.

\paragraph{Secrecy and covertness analysis} Let $\widehat{P}_{W_1W_2\mathbf{Z}}$ be the \modfirst{\ac{PMF} induced}  by the protocol and $\chi_2(\beta)\eqdef \beta \chi_2(Q_1^1\|Q_0^1) + (1-\beta)\chi_2(Q_1^0\|Q_0^0)$. By definition, we have \eqref{eq:sec-cov-an} on the top of next page,
\begin{figure*}[!t]

\begin{align}
\displaybreak[0]&S(\mathcal{C}|\mathbf{s}) + C(\mathcal{C}|\mathbf{s})\\
\displaybreak[0] &= \D{\widehat{P}_{W_1  W\modfirst{_2} \mathbf{Z}|\mathbf{s}}}{{P}_{W_1W\modfirst{_2}}^{\mathrm{unif}}\times \widehat{P}_{\mathbf{Z}|\mathbf{s}}} + \D{\widehat{P}_{\mathbf{Z}|\mathbf{s}}}{Q_0^{\proddist n}}\\
\displaybreak[0]&= \D{\widehat{P}_{W_1  W\modfirst{_2} \mathbf{Z}|\mathbf{s}}}{{P}_{W_1W\modfirst{_2}}^{\mathrm{unif}}\times \widehat{P}_{\mathbf{Z}|\mathbf{s}}} +\frac{1}{2} (\alpha_n + \kappa_n)^2\chi_2(\beta)n + O((\alpha_n + \kappa_n)^3n)\\
\displaybreak[0]&\stackrel{(a)}{=} \D{\widehat{P}_{W_1  W\modfirst{_2} \mathbf{Z}|\mathbf{s}}}{{P}_{W_1W\modfirst{_2}}^{\mathrm{unif}}\times \widehat{P}_{\mathbf{Z}|\mathbf{s}}} +\frac{1}{2} \alpha_n^2\chi_2(\beta)n + o(\alpha_n^2n)\\
\displaybreak[0]&\stackrel{(b)}{\leq} \V{\widehat{P}_{W_1  W\modfirst{_2} \mathbf{Z}|\mathbf{s}}\modfirst{,}{P}_{W_1W\modfirst{_2}}^{\mathrm{unif}}\times \widehat{P}_{\mathbf{Z}|\mathbf{s}}}\log(M_1M\modfirst{_2}) + \Hb{2 \V{\widehat{P}_{W_1  W\modfirst{_2} \mathbf{Z}|\mathbf{s}}\modfirst{,}{P}_{W_1W\modfirst{_2}}^{\mathrm{unif}}\times \widehat{P}_{\mathbf{Z}|\mathbf{s}}}}+\frac{1}{2} \alpha^2\chi_2(\beta)n+ o(\alpha_n^2n)\\
\displaybreak[0]&\stackrel{(c)}{\leq } \V{\widehat{P}_{W_1  W\modfirst{_2} \mathbf{Z}|\mathbf{s}}\modfirst{,}{P}_{W_1W\modfirst{_2}}^{\mathrm{unif}}\times \widehat{P}_{\mathbf{Z}|\mathbf{s}}}\left(O(n) + \log\frac{e}{2 \V{\widehat{P}_{W_1  W\modfirst{_2} \mathbf{Z}|\mathbf{s}}\modfirst{,}{P}_{W_1W\modfirst{_2}}^{\mathrm{unif}}\times \widehat{P}_{\mathbf{Z}|\mathbf{s}}} }\right)+\frac{1}{2} \alpha_n^2\chi_2(\beta)n + o(\alpha_n^2n).\label{eq:sec-cov-an}
\end{align}
\hrulefill
\end{figure*}
where $(a)$ follows since $\kappa_n = o(\alpha_n)$, $(b)$  follows from \cite[Problem 17.1]{csiszar2011information}, and $(c)$ follows from $\Hb{x} \leq x \log \frac{e}{x}$. To upper-bound $ \V{\widehat{P}_{W_1  W_3 \mathbf{Z}|\mathbf{s}}\modfirst{,}{P}_{W_1W_3}^{\mathrm{unif}}\times \widehat{P}_{\mathbf{Z}|\mathbf{s}}}$, let $\calE$ be the event $\{\mathbf{s}\notin \calS(M_1, M_2, M_3)\}$. By convexity of variational distance, we have
\begin{multline}
\V{\widehat{P}_{W_1  W_2 \mathbf{Z}|\mathbf{s}}\modfirst{,}{P}_{W_1W_2}^{\mathrm{unif}}\times \widehat{P}_{\mathbf{Z}|\mathbf{s}}}\\
\begin{split}
& \leq \V{\widehat{P}_{W_1  W\modfirst{_2} \mathbf{Z}|\calE\mathbf{s}}\modfirst{,}{P}_{W_1W\modfirst{_2}}^{\mathrm{unif}}\times \widehat{P}_{\mathbf{Z}|\calE\mathbf{s}}} \P{\calE} \\
&+ \V{\widehat{P}_{W_1  W\modfirst{_2} \mathbf{Z}|\calE^c\mathbf{s}}\modfirst{,}{P}_{W_1W\modfirst{_2}}^{\mathrm{unif}}\times \widehat{P}_{\mathbf{Z}|\calE^c\mathbf{s}}} \P{\calE^c}\\
&\leq n^{-4} + 2^{-\xi\kappa_n n' +1}\\
&\leq n^{-4} + 2^{-\omega(\log n)}\modfirst{.}
\end{split}
\end{multline}
Hence, for large enough $n$, we have $S(\mathcal{C}|\mathbf{s}) + C(\mathcal{C}|\mathbf{s}) \leq n^{-2} +\frac{1}{2} \alpha_n^2\chi_2(\beta)n+ o(\alpha_n^2n)$, which is vanishing.

\paragraph{Rate analysis} 
The covert rate of the protocol is 
\begin{align}
\frac{\log M_1}{\sqrt{nC(\calC|\mathbf{s})}} 
&\geq \frac{\log M_1}{\sqrt{n\pr{n^{-2} +\frac{1}{2} \alpha_n^2\chi_2(\beta)n+ o(\alpha_n^2n)}}}.
\end{align}
Moreover, by \eqref{eq:logM1-bound}, with probability \modfirst{at least} $2^{-\omega(\log n)}$, we have \eqref{eq:active-rate}. \modfirst{Finally, the required amount of secret common randomness for sharing $K$, sharing $\mathbf{J}$, and one-time-padding $\widehat{\beta}$ is $\log U = O(\log n)$, $n' \Hb{\kappa_n} = O(n' \kappa_n \log \frac{1}{\kappa_n}) =  o(n' \alpha_n)$, and $O(\log n)$, respectively. Since all three terms are $o(n' \alpha_n)$, and the amount of the generated key is $\Omega(n' \alpha_n)$,  the amount of secret common randomness is negligible.}
\begin{figure*}[!t]

\begin{align}
 \frac{\log M_1}{\sqrt{n\pr{n^{-2} +\frac{1}{2} \alpha_n^2\chi_2(\beta)n+ o(\alpha_n^2n)}}} 
& \geq  \frac{ (1 - \zeta) \pr{\avgI{{X};{Y}} - \zeta\alpha_n} n - (1+\zeta)\alpha_n\pr{ \beta {I}^1 + (1-\beta){I}^0 + \zeta }n - \zeta \alpha_n n}{\sqrt{n\pr{n^{-2} +\frac{1}{2} \alpha_n^2\chi_2(\beta)n+ o(\alpha_n^2n)}}} \\
  &=\sqrt{2}\frac{\D{(1-\beta)P_1^0+\beta P_1^1}{P_0} - \pr{(1-\beta)I^0+\beta I^1}}{\sqrt{\chi_2(\beta)}} -o(1) -O(\zeta).\label{eq:active-rate}
\end{align}
\hrulefill
\end{figure*}
\bibliographystyle{IEEEtran}
\bibliography{covert}

\appendices
\section{Proof of Lemma~\ref{lm:bernoulli_con}}
\label{sec:proof-bernoulli_con}
We first use the additivity of probability measures for disjoint events to split the probability into two parts, i.e.,
\begin{align}
&\P{\left|\frac{1}{1+\sum_{i=1}^nX_i } - \frac{1}{(n+1)p}\right| \geq\frac{\epsilon}{(n+1)p}} \\ \displaybreak[0]
&= \P{\frac{1}{1+\sum_{i=1}^nX_i } - \frac{1}{(n+1)p}\geq \frac{\epsilon}{(n+1)p}} \nonumber\\
&\phantom{===} + \P{\frac{1}{1+\sum_{i=1}^nX_i } - \frac{1}{(n+1)p} \leq  -\frac{\epsilon}{(n+1)p}} \\\displaybreak[0]
&= \P{\sum_{i=1}^nX_i \leq \frac{(n+1)p}{1+\epsilon} - 1}  \nonumber\\
&\phantom{============}+ \P{\sum_{i=1}^nX_i  \geq  \frac{(n+1)p}{1-\epsilon} - 1} \\\displaybreak[0]
&\leq  \P{\sum_{i=1}^nX_i \leq \frac{(n+1)p}{1+\epsilon} }  + \P{\sum_{i=1}^nX_i  \geq  \frac{np}{1-\epsilon} - 1} \displaybreak[0]\\
&= \P{\sum_{i=1}^nX_i \leq \pr{1- \frac{\epsilon - \frac{1}{n}}{1+\epsilon}}np}\nonumber \displaybreak[0]\\
&\phantom{=======}  + \P{\sum_{i=1}^nX_i  \geq  \pr{1+\frac{\epsilon}{1-\epsilon} - \frac{1}{np}} np}\displaybreak[0]\\\displaybreak[0]
&\stackrel{(a)}{\leq}  \P{\sum_{i=1}^nX_i \leq \pr{1- \frac{\epsilon}{2(1+\epsilon)}}np} \nonumber\displaybreak[0]\\
&\phantom{========} + \P{\sum_{i=1}^nX_i  \geq  \pr{1+\frac{\epsilon}{2(1-\epsilon)} } np}\\\displaybreak[0]
&\stackrel{(b)}{\leq}  \P{\sum_{i=1}^nX_i \leq \pr{1- \frac{\epsilon}{4}}np}\nonumber\\
& \phantom{===========} + \P{\sum_{i=1}^nX_i  \geq  \pr{1+\frac{\epsilon}{2} } np},
\end{align}
where $(a)$ follows since $\epsilon > \frac{2}{np}$, and $(b)$ follows since $\epsilon \in [0,1]$. To upper-bound the above terms, we use known Chernoff bounds \cite[Exercise 2.10]{boucheron2013concentration} stating that for $\mu\in]0, 1[$, we have
\begin{align}
\P{\sum_{i=1}^n X_i \leq (1-\mu)np} &\leq \exp\pr{-\frac{np\mu^2}{2}},\displaybreak[0]\\
\P{\sum_{i=1}^n X_i \geq (1+\mu)np} &\leq \exp\pr{-\frac{np\mu^2}{3}}.\displaybreak[0]
\end{align}
Therefore, we obtain
\begin{align}
\P{\sum_{i=1}^nX_i \leq \pr{1- \frac{\epsilon}{4}}np}  \leq  \exp\left(-\frac{np\epsilon^2}{32}\right),
\end{align}
and
\begin{align}
\P{\sum_{i=1}^nX_i \geq \pr{1+ \frac{\epsilon}{2}}np}
&\leq  \exp\left(-\frac{np\epsilon^2}{12}\right).
\end{align}
Combining these two inequalities completes the proof of \eqref{eq:concen_p}.

To prove \eqref{eq:concen_e}, we first define the event $\calE \eqdef\left\{\left|\frac{1}{1+\sum_{i=1}^nX_i } - \frac{1}{(n+1)p}\right| \geq\frac{\epsilon}{(n+1)p}\right\}$. By the law of total probability,
\begin{multline}
\E{\left|\frac{1}{1+\sum_{i=1}^nX_i} - \frac{1}{(n+1)p}\right| } \\
\begin{split}
&= \E{\left|\frac{1}{1+\sum_{i=1}^nX_i} - \frac{1}{(n+1)p}\right| \big | \calE}\P{\calE}\nonumber \\ 
&\phantom{====}+ \E{\left|\frac{1}{1+\sum_{i=1}^nX_i} - \frac{1}{(n+1)p}\right| \big | \calE^c}\P{\calE ^c}\\
&\leq \pr{1 + \frac{1}{(n+1)p}}e^{-\frac{np\epsilon^2}{32}} + \frac{\epsilon}{(n+1)p}.
\end{split}
\end{multline}
\section{Proof of Lemma~\ref{lm:oneshot_active_rel}}
\label{sec:proof-oneshot_active_rel}
By definition of our universal decoder and the construction of messages $W_1$ and $W_2$, we have
\begin{align}
&\P{W_1 \neq \widehat{W}_1}\displaybreak[0]\\
 &= \sum_{x,y, \widetilde{\mathbf{y}}, w_1, w_2} P_{XY}(x, y) Q_Y^{\proddist M_1M_2}(\widetilde{\mathbf{y}})P_{W_1W_2|Y\widetilde{\mathbf{Y}}}(w_1, w_2|y, \widetilde{\mathbf{y}})\nonumber\displaybreak[0]\\
 &\phantom{===}\times \indic{\exists w_1''\neq w_1:~\nu(x, \widetilde{y}_{w_1''w_2})\geq \nu(x, \widetilde{y}_{w_1w_2})}\displaybreak[0]\\
&\stackrel{(a)}{=}M_1M_2\sum_{x, y, \widetilde{\mathbf{y}}}P_{XY}(x, y) Q_Y^{\proddist M_1M_2}(\widetilde{\mathbf{y}})P_{W_1W_2|Y\widetilde{\mathbf{Y}}}(1, 1|y, \widetilde{\mathbf{y}})\nonumber\\
&\phantom{======} \times{\indic{\exists w_1''\neq 1:~\nu(x, \widetilde{y}_{w_1''1})\geq \nu(x, \widetilde{y}_{11})}}\\\displaybreak[0]
&\stackrel{(b)}{\leq} M_1M_2\sum_{x, y, \widetilde{\mathbf{y}}:\sum_{w_1'w_2'} \indic{y=\widetilde{y}_{w_1'w_2'}} \neq 0}P_{XY}(x, y) Q_Y^{\proddist M_1M_2}(\widetilde{\mathbf{y}})\nonumber\\\displaybreak[0]
&\phantom{=================}\times \frac{\indic{y=\widetilde{y}_{1 1}}}{\sum_{w_1'w_2'} \indic{y=\widetilde{y}_{w_1'w_2'}}}\nonumber\\\displaybreak[0]
&\phantom{=========}\times {\indic{\exists w_1''\neq 1:~\nu(x, \widetilde{y}_{w_1''1})\geq \nu(x, \widetilde{y}_{11})}}\nonumber\\\displaybreak[0]
&\phantom{===========}+ \sum_{y} P_Y(y) (1-Q_Y(y))^{M_1M_2}\\\displaybreak[0]
&\stackrel{(c)}{=}M_1M_2\sum_{x, \widetilde{\mathbf{y}}}P_{XY}(x, \widetilde{y}_{11}) Q_Y^{\proddist M_1M_2}(\widetilde{\mathbf{y}})\nonumber\\\displaybreak[0]
&\phantom{======= =}\times\frac{\indic{\exists w_1''\neq 1:~\nu(x, \widetilde{y}_{w_1''1})\geq \nu(x, \widetilde{y}_{11})}}{\sum_{w_1'w_2'} \indic{\widetilde{y}_{11}=\widetilde{y}_{w_1'w_2'}}}\nonumber\\\displaybreak[0]
&\phantom{===========}+ \sum_{y} P_Y(y) (1-Q_Y(y))^{M_1M_2},\label{eq:p_error-unbound}
\end{align}
where $(a)$ follows since  $Q_Y^{\proddist M_1M_2}$ is \ac{iid}, $(b)$ follows since the probability that for all $w_1'$ and $w_2'$, we have $Y\neq \widetilde{Y}_{w_1'w_2'}$ is $\sum_{y} P_Y(y) (1-Q_Y(y))^{M_1M_2}$, and $(c)$ follows since \modfirst{we can replace $y$ by $\widetilde{y}_{11}$ because of the term $\indic{y = \widetilde{y}_{11}}$}. We upper-bound the first term in \eqref{eq:p_error-unbound}  in two steps. Using Lemma~\ref{lm:bernoulli_con}, we first bound the \modfirst{difference between \eqref{eq:p_error-unbound} and the same expressions when replacing $\sum_{w_1'w_2'}\indic{\widetilde{y}_{11}=\widetilde{y}_{w_1'w_2'}}$ by its expected value $M_1M_2Q_Y(y)$. This follows from~\eqref{eq:sum-exp-pe} at} the top of next page,
\begin{figure*}[!t]
\begin{multline}
\left|\sum_{x, \widetilde{\mathbf{y}}} P_{XY}(x, \widetilde{y}_{11}) Q_Y^{\proddist M_1M_2}(\widetilde{\mathbf{y}})\indic{\exists w_1''\neq 1:~\nu(x, \widetilde{y}_{w_1''1})\geq \nu(x, \widetilde{y}_{11})}\pr{\frac{1}{M_1M_2Q_Y(\widetilde{y}_{11})}-\frac{1}{\sum_{w_1'w_2'}\indic{ \widetilde{y}_{11}=\widetilde{y}_{w_1'w_2'}}}}\right|\\ 
\begin{split}
&\leq\sum_{x, \widetilde{\mathbf{y}}} P_{XY}(x,  \widetilde{y}_{11}) Q_Y^{\proddist M_1M_2}(\widetilde{\mathbf{y}})\left|\frac{1}{M_1M_2Q_Y( \widetilde{y}_{11})}-\frac{1}{\sum_{w_1'w_2'}\indic{ \widetilde{y}_{11}=\widetilde{y}_{w_1'w_2'}}}\right|\\
&= \sum_{ \widetilde{\mathbf{y}}} P_{Y}( \widetilde{y}_{11}) Q_Y^{\proddist M_1M_2}(\widetilde{\mathbf{y}})\left|\frac{1}{M_1M_2Q_Y( \widetilde{y}_{11})}-\frac{1}{\sum_{w_1'w_2'}\indic{ \widetilde{y}_{11}=\widetilde{y}_{w_1'w_2'}}}\right|\\
&\stackrel{(a)}{\leq} \sum_{\widetilde{y}_{11}} P_Y(\widetilde{y}_{11}) Q_Y(\widetilde{y}_{11}) \left(\frac{\delta}{M_1M_2Q_Y(\widetilde{y}_{11})} + \pr{1 + \frac{1}{M_1M_2Q_Y(\widetilde{y}_{11})}}e^{-\frac{(M_1M_2-1)Q_Y(\widetilde{y}_{11})\delta^2}{32}}\right)\\
&=\sum_{\widetilde{y}_{11}} P_Y(\widetilde{y}_{11}) \left(\frac{\delta}{M_1M_2} + \pr{Q_Y(\widetilde{y}_{11}) + \frac{1}{M_1M_2}}e^{-\frac{(M_1M_2-1)Q_Y(\widetilde{y}_{11})\delta^2}{32}}\right)\\
&\leq  \frac{1}{M_1M_2}\pr{{\delta} + (1 + M_1M_2)e^{-\frac{(M_1M_2-1)\mu_Q\delta^2}{32}}}.
\end{split}\label{eq:sum-exp-pe}
\end{multline}
\hrulefill
\end{figure*}
where $(a)$ follows by applying Lemma~\ref{lm:bernoulli_con} when $\widetilde{y}_{11}$ is fixed and other components of $\mathbf{\widetilde{Y}}$ are \ac{iid} according to $Q_Y$. We now upper-bound
\begin{align*}
&\sum_{x, \widetilde{\mathbf{y}}}P_{XY}(x, \widetilde{y}_{11}) Q_Y^{\proddist M_1M_2}(\widetilde{\mathbf{y}})\displaybreak[0]\\
&\phantom{=====} \times \frac{\indic{\exists w_1''\neq 1:~\nu(x, y_{w_1''1})\geq \nu(x, y_{11})}}{Q_Y(\widetilde{y}_{11})}\displaybreak[0]\\
&= \sum_{x, \widetilde{y}_{11}}P_{XY}(x, \widetilde{y}_{11})\sum_{\widetilde{\mathbf{y}}\setminus\{\widetilde{y}_{11}\}} Q_Y^{\proddist M_1M_2}(\widetilde{\mathbf{y}})\displaybreak[0]\\
&\phantom{=====}\times\frac{\indic{\exists w_1''\neq 1:~\nu(x, y_{w_1''1})\geq \nu(x, y_{11})}}{Q_Y(\widetilde{y}_{11})}\displaybreak[0]\\
&\leq \sum_{x, \widetilde{y}_{11}}P_{XY}(x, \widetilde{y}_{11})\min\left(1,\right.\\
&\left. \sum_{\widetilde{\mathbf{y}}\setminus\{\widetilde{y}_{11}\}} Q_Y^{\proddist M_1M_2}(\widetilde{\mathbf{y}})\frac{\sum_{w_1'' \neq 1} \indic{~\nu(x, y_{w_1''1})\geq \nu(x, y_{11})}}{Q_Y(\widetilde{y}_{11})}\right)\\
&\leq  \sum_{x, y}P_{XY}(x, y)\min(1, M_1q(x, y)).
\end{align*}
Finally, to simplify our upper-bound on the average probability of error, we  bound the second term in \eqref{eq:p_error-unbound} as
\begin{align}
 \sum_{y} P_Y(y) (1-Q_Y(y))^{M_1M_2} 
  &\leq (1-\mu_Q)^{M_1M_2\displaybreak[0]}\\
   &= e^{\ln(1-\mu_Q) M_1M_2}\displaybreak[0]\\
 &\leq e^{- \frac{\mu_Q}{1-\mu_Q} M_1 M_2}\displaybreak[0]\\
 & \leq e^{- \frac{1}{32}\mu_Q M_1 M_2\delta^2},\label{eq:p-zero-dom}
\end{align}
which can be combined with \eqref{eq:sum-exp-pe} \modfirst{to obtain the desired result}.
\section{Proof of Lemma~\ref{lm:oneshot_active_sec}}
\label{sec:proof-oneshot_active_sec}
To simplify our notation, we treat ${P}_{W_1Z}$ as a random \ac{PMF} depending on $\mathbf{\widetilde{Y}}$, i.e.,
\begin{align}
{P}_{W_1Z}(w_1, z) \eqdef \sum_{y}  P_{YZ}(y, z)  \frac{\sum_{w_2}\indic{y=\widetilde{Y}_{w_1w_2}}}{\sum_{w_1'w_2'} \indic{y=\widetilde{Y}_{w_1'w_2'}}}
\end{align}
when $\sum_{w_1'w_2'} \indic{y=\widetilde{Y}_{w_1'w_2'}} \neq 0$ and ${P}_{W_1Z}(w_1, z) \eqdef P_Z(z) \frac{1}{M_1}$ otherwise. We can then write
\begin{multline}
\V{{P}_{W_1Z\widetilde{\mathbf{Y}}}, P_{W_1}^{\mathrm{unif}}\times{P}_{Z\widetilde{\mathbf{Y}}} } \\
= \E[\widetilde{\mathbf{Y}}]{\V{{P}_{W_1Z}, P_{W_1}^{\mathrm{unif}}\times{P}_{Z} }}.
\end{multline}
We first define $\overline{P}_{W_1Z}$  as
\begin{align}
\overline{P}_{W_1|Y}(w_1|y) &\eqdef \frac{1}{M_1M_2Q_Y(y)}\sum_{w_2}\indic{y = \widetilde{Y}_{w_1w_2}},\\
\overline{P}_{W_1Z}(w_1, z) &\eqdef \sum_{y}P_{YZ}(y, z) \overline{P}_{W_1|Y}(w_1|y),
\end{align}
which is not necessarily a \ac{PMF} because the sum over all $(w_1, z)$ may be less than one.
Note that
\begin{multline*}
\E[\mathbf{\widetilde{Y}}]{\|\overline{P}_{W_1Z}-{P}_{W_1Z}\|_1}\\
\begin{split}
 &= \sum_{w_1, z}\E[\mathbf{\widetilde{Y}}]{\left|\sum_{y}P_{YZ}(y, z) \left(\overline{P}_{W_1|Y}(w_1|y) - P_{W_1|Y}(w_1|y)\right)\right|}\\
& \leq \sum_{w_1, z, y}P_{YZ}(y, z)\E[\mathbf{\widetilde{Y}}]{\left| \overline{P}_{W_1|Y}(w_1|y) - P_{W_1|Y}(w_1|y)\right|}\\
& = \sum_{w_1, y}P_{Y}(y)\E[\mathbf{\widetilde{Y}}]{\left| \overline{P}_{W_1|Y}(w_1|y) - P_{W_1|Y}(w_1|y)\right|}\\
& \stackrel{(a)}{=} M_1  \sum_{y}P_{Y}(y)\E[\mathbf{\widetilde{Y}}]{\left| \overline{P}_{W_1|Y}(1|y) - P_{W_1|Y}(1|y)\right|},
\end{split}
\end{multline*}
where $(a)$ follows since $Q_Y^{\proddist M_1M_2}$ is \ac{iid}. We also have \eqref{eq:exp-cond-pw1} on the top of next page,
\begin{figure*}[!t]
\begin{multline}
\label{eq:exp-cond-pw1}
\E[\mathbf{\widetilde{Y}}]{\left| \overline{P}_{W_1|Y}(1|y) - P_{W_1|Y}(1|y)\right|} \\
\begin{split}
&\leq \sum_{w_2}\E[\mathbf{\widetilde{Y}}]{\indic{y=\widetilde{Y}_{1w_2}}\left|\frac{1}{M_1M_2Q_Y(y)}- \frac{1}{\sum_{w_1'w_2'} \indic{y=\widetilde{Y}_{w_1'w_2'}}}\right|\bigg| \sum_{w_1'w_2'} \indic{y=\widetilde{Y}_{w_1'w_2'}} \neq 0}\\ 
&\phantom{======================}\times \P[\mathbf{\widetilde{Y}}]{\sum_{w_1'w_2'}\indic{y=\widetilde{Y}_{w_1'w_2'}} \neq 0} + \frac{1}{M_1}\P[\mathbf{\widetilde{Y}}]{\sum_{w_1'w_2'}\indic{y=\widetilde{Y}_{w_1'w_2'}} = 0}\\
&= M_2 \E[\mathbf{\widetilde{Y}}]{\indic{y=\widetilde{Y}_{11}}\left|\frac{1}{M_1M_2Q_Y(y)}- \frac{1}{\sum_{w_1'w_2'} \indic{y=\widetilde{Y}_{w_1'w_2'}}}\right|\bigg| \sum_{w_1'w_2'} \indic{y=\widetilde{Y}_{w_1'w_2'}} \neq 0}\\ 
&\phantom{=======================}\times \P[\mathbf{\widetilde{Y}}]{\sum_{w_1'w_2'}\indic{y=\widetilde{Y}_{w_1'w_2'}} \neq 0} + \frac{1}{M_1}\P[\mathbf{\widetilde{Y}}]{\sum_{w_1'w_2'}\indic{y=\widetilde{Y}_{w_1'w_2'}} = 0}\\
&= M_2Q_Y(y) \E[\mathbf{\widetilde{Y}}\setminus\{\widetilde{Y}_{11}\} ]{\left|\frac{1}{M_1M_2Q_Y(y)}- \frac{1}{1+\sum_{w_1'w_2'\neq(1,1)} \indic{y=\widetilde{Y}_{w_1'w_2'}}}\right|}+ \frac{1}{M_1}\P[\mathbf{\widetilde{Y}}]{\sum_{w_1'w_2'}\indic{y=\widetilde{Y}_{w_1'w_2'}} = 0}\\
&= M_2Q_Y(y) \E[\mathbf{\widetilde{Y}}\setminus\{\widetilde{Y}_{11}\} ]{\left|\frac{1}{M_1M_2Q_Y(y)}- \frac{1}{1+\sum_{w_1'w_2'\neq(1,1)} \indic{y=\widetilde{Y}_{w_1'w_2'}}}\right|}+\frac{1}{M_1}(1-Q_Y(y))^{M_1M_2}\\
&\stackrel{(a)}{\leq} M_2Q_Y(y) \E[\mathbf{\widetilde{Y}}\setminus\{\widetilde{Y}_{11}\} ]{\left|\frac{1}{M_1M_2Q_Y(y)}- \frac{1}{1+\sum_{w_1'w_2'\neq(1,1)} \indic{y=\widetilde{Y}_{w_1'w_2'}}}\right|}+\frac{1}{M_1}e^{-\frac{1}{32}\mu_Q M_1M_2 \delta^2}.
\end{split}
\end{multline}
\hrulefill
\end{figure*}
where the derivation of $(a)$ is similar to that of \eqref{eq:p-zero-dom}. We can now use Lemma~\ref{lm:bernoulli_con}, for a particular $y$ and $\frac{2}{M_1M_2Q_Y(y)} < \delta < 1$ to obtain
\begin{align*}
&\E{\left|\frac{1}{M_1M_2Q_Y(y)}- \frac{1}{1+\sum_{w_1'w_2' \neq (1,1)} \indic{y=\widetilde{Y}_{w_1'w_2'}}}\right|}\displaybreak[0]\\
&\leq \frac{\delta}{M_1M_2Q_Y(y)} + \left(1+\frac{1}{M_1M_2Q_Y(y)}\right)e^{-\frac{(M_1M_2-1)Q_Y(y)\delta^2}{32}}\displaybreak[0]\\
&\leq \frac{\delta}{M_1M_2Q_Y(y)} + \left(1+\frac{1}{M_1M_2Q_Y(y)}\right)e^{-\frac{(M_1M_2-1)\mu_Q\delta^2}{32}}.
\end{align*}
Therefore, we obtain
\begin{multline}
\label{eq:app-error-sec}
\E[\mathbf{\widetilde{Y}}]{\|\overline{P}_{W_1Z}-{P}_{W_1Z}\|_1} 
\leq {\delta} + (1 + M_1M_2)e^{-\frac{(M_1M_2-1)\mu_Q\delta^2}{32}} \\+ \sum_{y}P_Y(y) (1-Q_Y(y))^{M_1M_2}.
\end{multline}

We next decompose $\overline{P}_{W_1Z}$ into two components  and define 
\begin{multline}
\overline{P}^1_{W_1Z}(w_1, z) \eqdef \sum_{y}P_{YZ}(y, z) \overline{P}_{W_1|Y}(w_1|y) \\
\times\indic{P_{YZ}(y, z) \geq \gamma P_Z(z) Q_Y(y)},
\end{multline}
\begin{multline}
\overline{P}^2_{W_1Z}(w_1, z) \eqdef \sum_{y}P_{YZ}(y, z) \overline{P}_{W_1|Y}(w_1|y)\\
\times \indic{P_{YZ}(y, z) < \gamma P_Z(z) Q_Y(y)},
\end{multline}
for which we upper-bound $\E{\|\overline{P}^1_{W_1Z} - \E{\overline{P}^1_{W_1Z}}\|_1} $ and $\E{\|\overline{P}^2_{W_1Z} - \E{\overline{P}^2_{W_1Z}}\|_1} $ as
\begin{align}
&\E{\|\overline{P}^1_{W_1Z} - \E{\overline{P}^1_{W_1Z}}\|_1}\\\displaybreak[0]
 &\leq \E{\|\overline{P}^1_{W_1Z}\|_1} + \|\E{\overline{P}^1_{W_1Z}}\|_1\\\displaybreak[0]
&=2\sum_{w_1, z}\E{\overline{P}^1(w_1, z)}\\\displaybreak[0]
&=2\sum_{w_1, z}\E{\sum_{y}P_{YZ}(y, z) \sum_{w_2}\frac{\indic{y=\widetilde{Y}_{w_1w_2}}}{M_1M_2Q_Y(y)}}\\\displaybreak[0]
&=2\sum_{y,z}P_{YZ}(y, z) \indic{P_{YZ}(y, z) \geq \gamma P_Z(z) Q_Y(y)},
\end{align}
and
\begin{align}
&\E{\|\overline{P}^2_{W_1Z} - \E{\overline{P}^2_{W_1Z}}\|_1}\\\displaybreak[0]
 &\leq \sum_{w_1, z}\E{\left|\overline{P}^2_{W_1Z}(w_1, z) -\E{\overline{P}^2(w_1, z)}\right|}\\\displaybreak[0]
&\leq \sum_{w_1, z}\sqrt{\Var{\overline{P}^2(w_1, z)}}\\\displaybreak[0]
&\leq \sum_{w_1, z}\sqrt{\sum_{y}\frac{P_{YZ}(y, z) ^2M_2Q_Y(y)}{M_1^2M_2^2Q_Y(y)^2}\modfirst{\indic{P_{YZ}(y, z) < \gamma P_Z(z) Q_Y(y)}} }\\\displaybreak[0]
&\leq\sum_{w_1, z}\sqrt{\sum_{y}\frac{P_{YZ}(y, z) \gamma  P_Z(z) Q_Y(y)M_2Q_Y(y)}{M_1^2M_2^2Q_Y(y)^2} }\\\displaybreak[0]
&= \sqrt{\frac{\gamma}{M_2}}.
\end{align}
Therefore, by the triangle inequality, we obtain
\begin{multline}
\label{eq:app-sec}
\E{\|\overline{P}_{W_1Z} - \E{\overline{P}_{W_1Z}}\|_1}\\ \leq 2\sum_{y,z}P_{YZ}(y, z) \indic{P_{YZ}(y, z) \geq \gamma P_Z(z) Q_Y(y)}+ \sqrt{\frac{\gamma}{M_2}}.
\end{multline}
Combining \eqref{eq:app-error-sec} and \eqref{eq:app-sec} and noting that $\E{\overline{P}_{W_1Z}} = P_{W_1}^{\mathrm{unif}}\times{P}_{Z}$  \modfirst{completes} the proof.
\section{Proof of LEmma~\ref{lm:active_random_code}}
\label{sec:proof-active_random_code}
Applying  Lemma~\ref{lm:oneshot_active_rel} to $(W_1, W_3)$ and $W_2$ for a fixed $\mathbf{s}$, we obtain
\begin{align}
&\E[F, \Phi]{P_e(F, \Phi|\mathbf{s})} \displaybreak[0]\\
&\leq \sum_{\mathbf{x}, \mathbf{y}}Q_{X}^{\pn}(\mathbf{x})W_{Y|XS}^\pn(\mathbf{y}|\mathbf{x}\mathbf{s})\min(1, M_{1}M_{3}q(\mathbf{x}, \mathbf{y}))\nonumber \displaybreak[0]\\
&\phantom{======}+ {\delta}+ (2 + M_{1}M_{2}M_{3})e^{-\frac{(M_{1}M_{2}M_{3}-1)\mu_0^n\delta^2}{32}}\displaybreak[0] \\
&= \sum_{\mathbf{x}, \mathbf{y}}Q_{XY|S}^{\pn}(\mathbf{x},\mathbf{y}|\mathbf{s})\min(1, M_{1}M_{3}q(\mathbf{x}, \mathbf{y}))\nonumber\displaybreak[0]\\
&\phantom{=====} + {\delta}+ (2 + M_{1}M_{2}M_{3})e^{-\frac{\modfirst{(}M_{1}M_{2}M_{3}-1)\mu_0^n\delta^2}{32}},\label{eq:rel-universal-aux}
\end{align}
where
\begin{align}
q(\mathbf{x}, \mathbf{y}) \eqdef \sum_{\widetilde{\mathbf{y}}} P_0^\pn(\widetilde{\mathbf{y}})\indic{I(\mathbf{x}\wedge \mathbf{y}) \leq I(\mathbf{x}\wedge \widetilde{\mathbf{y}}) }.
\end{align}
To upper-bound $q(\mathbf{x}, \mathbf{y})$, let $V_X$ and $V_{Y|X}$ be the type of $\mathbf{x}$  and the conditional type of $\mathbf{y}$ given $\mathbf{x}$, respectively. Then, we have
\begin{align*}
&\sum_{\widetilde{\mathbf{y}}} P_0^\pn(\widetilde{\mathbf{y}})\indic{I(\mathbf{x}\wedge \mathbf{y}) \leq I(\mathbf{x}\wedge \widetilde{\mathbf{y}}) }\displaybreak[0] \\
&= \sum_{\widetilde{V}_{Y|X}} P_0^\pn(\calT_{\widetilde{V}_{Y|X}}(\mathbf{x}))\displaybreak[0]\\
&\phantom{===========}\times \indic{I(V_X, \widetilde{V}_{Y|X}) \geq I(V_X, V_{Y|X}) }\displaybreak[0]\\
&\stackrel{(a)}{\leq} \sum_{\widetilde{V}_{Y|X} \in\calP_n(\calY|\calX)}  2^{-n \D{\widetilde{V}_{Y|X}}{P_0|V_X}}  \\ &\phantom{===========}\times \indic{I(V_X, \widetilde{V}_{Y|X}) \geq I(V_X, V_{Y|X}) }\displaybreak[0]\\
&\leq (n+1)^{|\calX||\calY|}\displaybreak[0]\\
&\times 2^{-n\mathop{\min}_{\widetilde{V}_{Y|X}:I(V_X, \widetilde{V}_{Y|X}) \geq I(V_X, V_{Y|X})} \D{\widetilde{V}_{Y|X}}{P_0|V_X}}\displaybreak[0]\\
&=(n+1)^{|\calX||\calY|}\displaybreak[0]\\
&\times 2^{-n\min_{\widetilde{V}_{Y|X}:I(V_X, \widetilde{V}_{Y|X}) \geq I(V_X, V_{Y|X})} I(V_X, \widetilde{V}_{Y|X}) + \D{\widetilde{V}_{Y|X}\circ V_X}{P_0}}\displaybreak[0]\\
&\leq(n+1)^{|\calX||\calY|}2^{-n\min_{\widetilde{V}_{Y|X}:I(V_X, \widetilde{V}_{Y|X}) \geq I(V_X, V_{Y|X})} I(V_X, \widetilde{V}_{Y|X}) }\displaybreak[0]\\
&= (n+1)^{|\calX||\calY|} 2^{-nI(V_X, V_{Y|X})}\displaybreak[0]\\
&=(n+1)^{|\calX||\calY|} 2^{-nI({V}_{X|S}\circ Q_S, {V}_{Y|XS}\circ Q_S)},
\end{align*}
where $(a)$ follows from \cite[Lemma 2.6]{csiszar2011information}. Substituting the above upper-bound in the first term of the right hand side of \eqref{eq:rel-universal-aux} and using $Q_{XY|S}^{\pn}(\calT_{V_{XY|S}}(\mathbf{s})|\mathbf{s}) \leq 2^{-n\D{V_{XY|S}}{Q_{XY|S}|Q_S}}$ \cite[Equation (2.8)]{csiszar2011information}, we have \eqref{eq:type-bound-error} on the top of next page,
\begin{figure*}[!t]

\begin{align}
&\sum_{\mathbf{x}, \mathbf{y}}Q_{X}^{\pn}(\mathbf{x})W_{Y|XS}^\pn(\mathbf{y}|\mathbf{x}\mathbf{s})\min(1, M_{1}M_{3}q(\mathbf{x}, \mathbf{y}))\\
&~~~~~~\leq \sum_{{V}_{XY|S}\in\calP_n(\calX\times \calY|\calS)} 2^{-n\D{{V}_{XY|S}}{Q_{XY|S}|Q_S}}\min\pr{1, M_{1}M_{3}(n+1)^{|\calX||\calY|}2^{-nI({V}_{X|S}\circ Q_S, {V}_{Y|XS}\circ Q_S)}}\\
&~~~~~~= \sum_{{V}_{XY|S}\in\calP_n(\calX\times \calY|\calS)} 2^{-n\D{{V}_{XY|S}}{Q_{XY|S}|Q_S}}2^{-n[I({V}_{X|S}\circ Q_S, {V}_{Y|XS}\circ Q_S) - \log M_{1}M_{3}/n - O(\log n)/n]^+}\\
&~~~~~~\leq (n+1)^{2|\calX||\calY|}2^{-n\min_{{V}_{XY|S}} \pr{\D{{V}_{XY|S}}{Q_{XY|S}|Q_S} +  [I({V}_{X|S}\circ Q_S, {V}_{Y|XS}\circ Q_S) - \log M_{1}M_{3}/n - O(\log n)/n]^+}}\\
&~~~~~~\leq 2^{-n\min_{{V}_{XY|S}} \pr{\D{{V}_{XY|S}}{Q_{XY|S}|Q_S} +  [I({V}_{X|S}\circ Q_S, {V}_{Y|XS}\circ Q_S) - \log M_{1}M_{3}/n ]^+- O(\log n)/n}}\\
&~~~~~~\stackrel{(a)}{\leq} 2^{-n\min_{{V}_{XY}} \pr{\D{{V}_{XY}}{Q_{XY}} +  [I({V}_{X}, {V}_{Y|X}) - \log M_{1}M_{3}/n ]^+- O(\log n)/n}}\\
&~~~~~~\stackrel{(b)}{\leq} 2^{-n\min_{{V}_{XY}} \pr{\D{{V}_{XY}}{Q_{XY}} +  [I({V}_{X}, {V}_{Y|X}) - (1-\zeta)I(Q_X, Q_{Y|X}) ]^+- O(\log n)/n}}.\label{eq:type-bound-error}
 \end{align}
 \hrulefill
\end{figure*}
 where $(a)$ follows because \modfirst{$\D{{V}_{XY|S}}{Q_{XY|S}|Q_S} \geq \D{{V}_{XY}}{Q_{XY}}$ by convexity of the KL-divergence}, and $(b)$ follows from \eqref{eq:M_13_bound}. We next state a result that shows that, for all $V_{XY}$,  $\D{{V}_{XY}}{Q_{XY}} $ and $[I({V}_{X}, {V}_{Y|X}) - (1-\zeta)I(Q_X, Q_{Y|X}) ]^+$ cannot be \modfirst{simultaneously small}.
 
 \begin{lemma}
 \label{lm:exponent-rel}
 For a \ac{PMF} $V_{XY}$ with $\D{{V}_{XY}}{Q_{XY}} \leq \epsilon$, we have
 \begin{multline*}
 I({V}_{X}, {V}_{Y|X}) \geq  \pr{\alpha - \sqrt{2\epsilon \alpha}} \left( \D{Q_{Y|X=1}}{Q_{Y|X=0}}\right. \\
 \left.- \sqrt{\frac{\epsilon}{\alpha - \sqrt{2\epsilon\alpha}  }} \pr{B +\frac{1}{2}\log \frac{1}{\sqrt{\frac{\epsilon}{\alpha - \sqrt{2\epsilon\alpha} }}}}\right) - \frac{\pr{\alpha + \sqrt{2\epsilon}}^2 \card{\calY}}{ \mu_0 - \sqrt{\frac{\epsilon}{1-\widetilde{\alpha}}}},
 \end{multline*}
 where $\alpha \eqdef Q_X(1)$, $\mu_0\eqdef \min_{y} Q_{Y|X}(y|0)$, and $B$ is a constant that depends only on $\card{\calY}$ and $\mu_0$.
 \end{lemma}
 \begin{IEEEproof}
 See Appendix~\ref{sec:proof-exponent-rel}.
 \end{IEEEproof}
 To lower-bound the exponent in \eqref{eq:type-bound-error},
 \begin{multline}
 n\min_{{V}_{XY}} (\D{{V}_{XY}}{Q_{XY}} +  [I({V}_{X}, {V}_{Y|X}) \\ - (1-\zeta)I(Q_X, Q_{Y|X}) ]^+- \modfirst{O\pr{\log n}/n}),
 \end{multline}
 we consider two cases for $V_{XY}$. For $\{\epsilon_n\}_{n\geq1} = \omega\pr{\frac{\log n}{n}}\cap o(\alpha_n)$ and $\D{V_{XY}}{Q_{XY}}\geq \epsilon_n$, we have
 \begin{multline}
 \D{{V}_{XY}}{Q_{XY}} +  [I({V}_{X}, {V}_{Y|X}) - (1-\zeta)I(Q_X, Q_{Y|X}) ]^+\\
 - O(\log n)/n
\geq \epsilon_n + O\pr{\frac{\log n}{n}} 
\stackrel{(a)}{=} \omega\pr{\frac{\log n}{n}},
 \end{multline}
 where $(a)$ follows since $\epsilon_n = \omega(\log n/n)$. For the case when $\D{V_{XY}}{Q_{XY}}\leq \epsilon_n$, applying Lemma~\ref{lm:exponent-rel}, we obtain
 \begin{align}
  &\D{{V}_{XY}}{Q_{XY}} +  [I({V}_{X}, {V}_{Y|X})\nonumber \\\displaybreak[0]
  &\phantom{=======}- (1-\zeta)I(Q_X, Q_{Y|X}) ]^+- O(\log n/n)\\\displaybreak[0]
  &\geq  [I({V}_{X}, {V}_{Y|X}) - (1-\zeta)I(Q_X, Q_{Y|X}) ]^+- O(\log n/n)\\\displaybreak[0]
  &\geq  \left[\pr{\alpha_n - \sqrt{2\epsilon_n \alpha_n}} \left(\D{Q_{Y|X=1}}{Q_{Y|X=0}}\right.\right.\nonumber\\\displaybreak[0]
 & \left.\left. - \sqrt{\frac{\epsilon_n}{\alpha_n - \sqrt{2\epsilon_n\alpha_n}  }} \pr{B +\frac{1}{2}\log \frac{1}{\sqrt{\frac{\epsilon_n}{\alpha_n - \sqrt{2\epsilon_n\alpha_n} }}}}\right)\right. \nonumber\\\displaybreak[0]
  &\phantom{} \left. -\frac{\pr{\alpha_n + \sqrt{2\epsilon_n}}^2 \card{\calY}}{ \mu_0 - \sqrt{\frac{\epsilon_n}{1-\widetilde{\alpha_n}}}}- (1-\zeta)I(Q_X, Q_{Y|X}) \right]^+\\\displaybreak[0]
  &\phantom{=====================}- O\pr{{\log n}/{n}}\\
  &\stackrel{(a)}{=} [\alpha_n(1 - o(1))\pr{\D{Q_{Y|X=1}}{Q_{Y|X=0}} - o(1)} - o(\alpha_n) \nonumber \\\displaybreak[0]
  &\phantom{=======}-(1-\zeta)I(Q_X, Q_{Y|X}) ]^+- O(\log n/n)\\\displaybreak[0]
  &=[\alpha_n\D{Q_{Y|X=1}}{Q_{Y|X=0}}    -(1-\zeta)I(Q_X, Q_{Y|X})\nonumber\\\displaybreak[0]
  &\phantom{==============}- o(\alpha_n) ]^+- O(\log n/n)\\
  &\stackrel{(b)}{\geq}[\alpha_n\D{Q_{Y|X=1}}{Q_{Y|X=0}}  -(1-\zeta)\alpha_n\nonumber\\\displaybreak[0]
  &\phantom{==}\times\D{Q_{Y|X=1}}{Q_{Y|X=0}}- o(\alpha_n) ]^+- O(\log n/n)\\
  &= (1-o(1))\zeta \alpha_n \D{Q_{Y|X=1}}{Q_{Y|X=0}}\\
  &= \omega\pr{\frac{\log n}{n}},
 \end{align}
 where $(a)$ follows since $\sqrt{2\epsilon_n\alpha_n} = o(\alpha_n)$, $ \sqrt{\frac{\epsilon_n}{\alpha_n - \sqrt{2\epsilon_n\alpha_n}  }} = o(1)$, and $\pr{\alpha_n + \sqrt{2\epsilon_n}}^2 = o(\alpha_n)$, and $(b)$ follows from \cite[Lemma 1]{Bloch2016a}.
Therefore, we conclude that
\begin{multline}
 n\min_{{V}_{XY}} (\D{{V}_{XY}}{Q_{XY}} +  [I({V}_{X}, {V}_{Y|X}) \\
 - (1-\zeta)I(Q_X, Q_{Y|X}) ]^+- O(\log n)/n) = \omega(\log n),
\end{multline}
and
\begin{align}
\sum_{\mathbf{x}, \mathbf{y}}Q_{X}^{\pn}(\mathbf{x})W_{Y|XS}^\pn(\mathbf{y}|\mathbf{x}\mathbf{s})\min(1, M_{1}M_{3}q(\mathbf{x}, \mathbf{y})) \leq 2^{-\omega(\log n)}.
\end{align}

 Furthermore, \modfirst{the choice} $\delta = 2^{-\frac{1}{3}\zeta n\log \frac{1}{\mu_0}  }$ satisfies $\frac{2}{M_1M_2M_3 \mu_0^n} \leq \delta < 1$ for large $n$ and we obtain
\begin{multline}
\delta + (\modfirst{2}+M_1M_2M_3)e^{-\frac{(M_1M_2M_3-1)\mu_0\delta^2}{32}} \\
\leq \frac{ 2^{-\frac{1}{3}\zeta \log \frac{1}{\mu_0} n }}{1- 2^{-\frac{1}{3}\zeta \log \frac{1}{\mu_0} n }} + \pr{\modfirst{2}+2^{\lceil (1+\zeta)n \log \frac{1}{\mu_0}\rceil}}e^{-\frac{2^{\frac{1}{2}\zeta \log \frac{1}{\mu_0} n}}{32}},
\end{multline}
which is less than $e^{-\zeta' n}$ for large enough $n$ and $\zeta' > 0$ independent of $n$. 

To analyze the secrecy,  we fix $\mathbf{s} \in \calS^n$ with $\wt{\mathbf{s}} = \beta n$ and define \modfirst{the} \ac{PMF}  $P_{\mathbf{Y}\mathbf{Z}}(\mathbf{y}, \mathbf{z}) \eqdef \sum_{\mathbf{x}}Q_X^\pn(\mathbf{x}) W_{YZ|XS}^\pn(\mathbf{y}\mathbf{z}|\mathbf{x}\mathbf{s})$, by Lemma~\ref{lm:oneshot_active_sec}, we have 
 \begin{multline*}
\E[F,\Phi]{ S\pr{F, \Phi|\mathbf{s}} }\leq \P[P_{\mathbf{Y}\mathbf{Z}}]{\sum_{i=1}^n\log \frac{P_{Y_i|Z_i}(Y_i|Z_i)}{P_0(Y_i)} \geq \log \gamma}\\
 +\frac{1}{2}\sqrt{\frac{\gamma}{M_3}} + \frac{1}{2}\delta + \frac{1}{2}(\modfirst{2} + M_1M_2M_3)e^{-\frac{(M_1M_2M_3-1)\mu_0^n\delta^2}{32}}.
 \end{multline*}
 Moreover, for $t \eqdef \gamma - \sum_{i=1}^n \E{\log \frac{P_{Y_i|Z_i}(Y_i|Z_i)}{P_0(Y_i)}} \geq 0$ and 
\begin{multline} 
 C \eqdef \max_{y, z}\left(\log \frac{\sum_{x}Q_X(x) W_{YZ|XS}(y,z|x,0))}{P_0(y)}\right.
 \\\left.,\log \frac{\sum_{x}Q_X(x) W_{YZ|XS}(y,z|x,\modfirst{1}))}{P_0(y)}\right) = O(1),
\end{multline} 
 Bernstein's inequality yields that for 
 \begin{multline}
  \P[P_{\mathbf{Y}\mathbf{Z}}]{\sum_{i=1}^n\log \frac{P_{Y_i|Z_i}(Y_i|Z_i)}{P_0(Y_i)} \geq \log \gamma}\\
   \leq \exp\pr{\frac{t^2}{\sum_{i=1}^n\Var{\log \frac{P_{Y_i|Z_i}(Y_i|Z_i)}{P_0(Y_i)}}+ \frac{1}{3}Ct} }.
 \end{multline}
 By \cite[Lemma~2]{Tahmasbi2017}, we also have $\E[P_{Y_iZ_i}]{\log \frac{P_{Y_i|Z_i}(Y_i|Z_i)}{P_0(Y_i)}} = \avgI{Y_i; Z_i} + \D{P_{Y_i}}{P_0} = \alpha_n I^{s_i} + O(\alpha_n^2)$, where $I^s$ is defined in the statement of Theorem~\ref{th:main_active}. Thus, by choosing $ \gamma =  (1+\zeta/2) \alpha_n \pr{\beta I^1 + (1-\beta)I^0}$, we obtain that $\E[F,\Phi]{ S\pr{F, \Phi|\mathbf{s}} }\leq 2^{-\xi n}$ for some $\xi > 0$ small enough.
\section{Proof of Lemma~\ref{lm:exponent-rel}}
\label{sec:proof-exponent-rel}
For a fixed $V_{XY}$ and $Q_{XY}$, we first define
\begin{align}
\alpha &\eqdef Q_X(1),\quad \widetilde{\alpha} &\eqdef V_X(1),\\
P_0 &\eqdef Q_{Y|X=0},\quad \widetilde{P}_0 &\eqdef V_{Y|X=0},\\
P_1 &\eqdef Q_{Y|X=1},\quad \widetilde{P}_1 &\eqdef V_{Y|X=1},\\
P_{s} &\eqdef sQ_{Y|X=1}+(1-s)Q_{Y|X=0},\\
\widetilde{P}_s &\eqdef sV_{Y|X=1}+(1-s)V_{Y|X=0}.
\end{align}
By \cite[Lemma 1]{Bloch2016a}, we have
\begin{align}
I(V_{XY}) 
&= \widetilde{\alpha}\D{\widetilde{P}_1}{\widetilde{P}_0} - \D{\widetilde{P}_{\widetilde{\alpha}}}{\widetilde{P}_0}\\
&\geq \widetilde{\alpha}\D{\widetilde{P}_1}{\widetilde{P}_0} - \widetilde{\alpha}^2 \chi_2(\widetilde{P}_1\|\widetilde{P}_0).
\end{align}
Moreover, by the chain rule for relative entropy, we can write $\D{V_{XY}}{Q_{XY}}$ as
\begin{multline}
\D{V_{XY}}{Q_{XY}} \\
\begin{split}
&= \D{V_{X}}{Q_{X}} + \D{V_{Y|X}}{Q_{Y|X}|V_X}\\
&= \D{\widetilde{\alpha}}{\alpha} + \widetilde{\alpha} \D{\widetilde{P}_1}{P_1} + (1-\widetilde{\alpha}) \D{\widetilde{P}_0}{P_0},\label{eq:d-v-q}
\end{split}
\end{multline}
where $\D{p}{q} \eqdef p\log(p/q) + (1-p)\log((1-p)/(1-q))$. Since all terms in terms in \eqref{eq:d-v-q} are positive, our assumption that $\D{V_{XY}}{Q_{XY}} \leq \epsilon$ implies that
\begin{align}
\D{\widetilde{\alpha}}{\alpha} &\leq \epsilon,\\
\D{\widetilde{P}_1}{P_1} &\leq \frac{\epsilon}{\widetilde{\alpha} },\\
\D{\widetilde{P}_0}{P_0} &\leq \frac{\epsilon}{1-\widetilde{\alpha}}.
\end{align}
Using the inequalities $\D{p}{q} \geq (p-q)^2/(2q)$ for $p\leq q$ and $\D{p}{q} \geq (p-q)^2/(2p)$ for $q\leq p$, we obtain
\begin{align}
\alpha - \sqrt{2\epsilon\alpha} \leq \widetilde{\alpha} \leq \alpha + \sqrt{2\epsilon}.
\end{align}
Furthermore, Pinsker's inequality yields that $\V{\widetilde{P}_1,P_1} \leq \sqrt{\frac{\epsilon}{\widetilde{\alpha} }}$ and $
\V{\widetilde{P}_0,P_0} \leq \sqrt{\frac{\epsilon}{1-\widetilde{\alpha}}}$. Hence,
\begin{align}
&\D{\widetilde{P}_1}{\widetilde{P}_0} \\\displaybreak[0]
&=  \D{{P}_1}{{P}_0} + \D{\widetilde{P}_1}{\widetilde{P}_0}  - \D{{P}_1}{{P}_0}  \\\displaybreak[0]
&\geq   \D{{P}_1}{{P}_0} - \sqrt{\frac{\epsilon}{\alpha - \sqrt{2\epsilon\alpha} }}\nonumber\\\displaybreak[0]
&\times \pr{\frac{ \log(\card{\calY}-1)}{2}  + \card{\calY}\log \frac{1}{\mu_0} + \frac{\card{\calY}^2}{\mu_0-  \sqrt{\frac{\epsilon}{\alpha - \sqrt{2\epsilon\alpha} }}}}\nonumber\\\displaybreak[0]
&\phantom{================} -\Hb{\frac{ \sqrt{\frac{\epsilon}{\alpha - \sqrt{2\epsilon\alpha} }}}{2}}\\
&\geq \D{{P}_1}{{P}_0} - \sqrt{\frac{\epsilon}{\alpha - \sqrt{2\epsilon\alpha}  }} \left(\frac{ \log(\card{\calY}-1)}{2}  + \card{\calY}\right.\nonumber\\\displaybreak[0]
&\left. \times \log \frac{1}{\mu_0} + \frac{\card{\calY}^2}{\mu_0-  \sqrt{\frac{\epsilon}{\alpha - \sqrt{2\epsilon\alpha}  }}} +\frac{1}{2}\log \frac{2e}{\sqrt{\frac{\epsilon}{\alpha - \sqrt{2\epsilon\alpha} }}}\right)\\
&= \D{{P}_1}{{P}_0} - \sqrt{\frac{\epsilon}{\alpha - \sqrt{2\epsilon\alpha}  }} \pr{B +\frac{1}{2}\log \frac{1}{\sqrt{\frac{\epsilon}{\alpha - \sqrt{2\epsilon\alpha} }}}},
\end{align}
where $B$  depends only on $\mu_0$ and $\card{\calY}$. Also because $\chi_2(\widetilde{P}_1\|\widetilde{P}_0) \leq \card{\calY}\frac{1}{\widetilde{\mu}_0}$ and $\widetilde{\mu}_0 \geq \mu_0 - \sqrt{\frac{\epsilon}{1-\widetilde{\alpha}}}$, we have
\begin{align}
I(V_{XY}) 
&= \widetilde{\alpha}\D{\widetilde{P}_1}{\widetilde{P}_0} - \D{\widetilde{P}_{\widetilde{\alpha}}}{\widetilde{P}_0}\\\displaybreak[0]
&\geq \widetilde{\alpha}\D{\widetilde{P}_1}{\widetilde{P}_0} - \widetilde{\alpha}^2 \chi_2(\widetilde{P}_1\|\widetilde{P}_0)\\\displaybreak[0]
&\geq \pr{\alpha - \sqrt{2\epsilon \alpha}} \left(\D{{P}_1}{{P}_0} - \sqrt{\frac{\epsilon}{\alpha - \sqrt{2\epsilon\alpha}  }}\right.\nonumber\\\displaybreak[0]
&\left.\phantom{=}\times \pr{B +\frac{1}{2}\log \frac{1}{\sqrt{\frac{\epsilon}{\alpha - \sqrt{2\epsilon\alpha} }}}}\right) - \pr{\alpha + \sqrt{2\epsilon}}^2\nonumber\\\displaybreak[0]
&\phantom{=============}\times \card{\calY}\frac{1}{ \mu_0 - \sqrt{\frac{\epsilon}{1-\widetilde{\alpha}}}}\\
&= \alpha \D{P_1}{P_0} - \sqrt{2\epsilon\alpha}\D{P_1}{P_0}\nonumber \\\displaybreak[0]
&-\sqrt{\epsilon\pr{\alpha-\sqrt{2\epsilon\alpha} }} \pr{B +\frac{1}{2}\log \frac{1}{\sqrt{\frac{\epsilon}{\alpha - \sqrt{2\epsilon\alpha} }}}}\nonumber\\
&\phantom{=====} - \pr{\alpha + \sqrt{2\epsilon}}^2 \card{\calY}\frac{1}{ \mu_0 - \sqrt{\frac{\epsilon}{1-\widetilde{\alpha}}}}.
\end{align}
\end{document}